\global\def\draftcontrol{0}

   \def\versionno{ normal functions in CICY }

\catcode`\@=11

\expandafter\ifx\csname draftcontrol\endcsname\relax\global\def\draftcontrol{0} 
\fi 

{\count255=\time\divide\count255 by 60 
\xdef\hourmin{\number\count255} 
\multiply\count255 by-60\advance\count255 by\time 
\xdef\hourmin{\hourmin:\ifnum\count255<10 0\fi\the\count255}} 
\def\draftdate{\number\month/\number\day/\number\year\ \ \ \hourmin } 

\newcommand\makepapertitle{\par

  \begingroup 
    \renewcommand\thefootnote{\@fnsymbol\c@footnote}%
    \def\@makefnmark{\rlap{\@textsuperscript{\normalfont\@thefnmark}}}%
    \long\def\@makefntext##1{\parindent 1em\noindent 
            \hb@xt@1.8em{%
                \hss\@textsuperscript{\normalfont\@thefnmark}}##1}%
     \newpage 
     \global\@topnum\z@   
     \@makepapertitle 
     \thispagestyle{empty}\@thanks 
  \endgroup 
  \setcounter{footnote}{0}%
  \global\let\thanks\relax 
  \global\let\makepapertitle\relax 
  \global\let\@makepapertitle\relax 
  \global\let\@thanks\@empty 
  \global\let\@author\@empty 
  \global\let\@date\@empty 
  \global\let\@title\@empty 
  \global\let\title\relax 
  \global\let\author\relax 
  \global\let\date\relax 
  \global\let\and\relax 
  \def\version{\let\version\@version\@gobble} 
} 
\def\@makepapertitle{%
  \newpage 
   \ifnum\draftcontrol=1 {} 
   \version\versionno 
   \vskip 5.5em%
   \else 
   \hfill\hbox to 3.5cm {\parbox{5cm}{\@pubnum}\hss}%
   \vskip 6.5em%
   \fi 
   \begin{center}%
   \let \footnote \thanks 
      {\hskip -0\textwidth \hbox to 1\textwidth%
        {\centerline{\Large\bf{\noindent\@title}}}}%
     \vskip 2em%
     {\normalsize
       \lineskip .5em%
       \begin{tabular}[t]{c}%
         \@author 
       \end{tabular}\par}%
     \vskip 1.5em%
     {\@bstract}%
     \end{center}%
     \vfill
     \@date%
     \vskip 1.5em%
   \par 
} 

\gdef\@pubnum{} 
\def\pubnum#1{%
  \gdef\@pubnum{#1}} 

\gdef\@bstract{} 
\def\Abstract#1{%
  \gdef\@bstract{%
   \parbox{\textwidth-0pc}{%
   \centerline{\bf Abstract}\penalty1000 
   \noindent
   \renewcommand\baselinestretch{1.0} 
   {#1}}} 
} 

\gdef\@email{}
\def\email#1{%
   \gdef\@email{%
   Email: {\tt #1}}
}

\def\ps@paper{\let\@mkboth\@gobbletwo%
     \ifnum\draftcontrol=1 
        \def\@oddfoot{\hbox to \textwidth{\tiny \versionno \hfil\tiny\draftdate}%
        \hskip -\textwidth \hbox to \textwidth{\hfil\rm\thepage\hfil}}%
     \else\def\@oddfoot{\hbox to \textwidth{\hfil\rm\thepage\hfil}} 
     \fi 
     \let\@evenfoot\@oddfoot 
} 

\def\body{\clearpage 
          \pagestyle{paper} 
        } 
\newenvironment{acknowledgments}{%
\vskip 3.25ex 
\addcontentsline{toc}{section}{Acknowledgments}
\noindent {\bf Acknowledgments} 
} 

\def\@version#1{\ifnum\draftcontrol=1 
\typeout{}\typeout{#1}\typeout{} 
\vskip3mm\centerline{\hbox{\fbox{\normalsize{\tt DRAFT -- #1 -- } 
                   {\draftdate}}}}\vskip3mm 
\fi} 
\let\version\@version 
\long\def\eqlabel#1{\ifnum\draftcontrol=1 
                    \tag@false  
                    \tag*{(\theequation) \hbox to -0.2cm{\hspace{0cm}\small{#1}\hss}} 
                    \refstepcounter{equation}  
                    \edef\@currentlabel{\theequation} 
                    \ltx@label{#1}          
                    \else 
                    \label{#1} 
                    \fi 
                    } 
\let\st@bibitem\@bibitem 
\let\st@lbibitem\@lbibitem 
\ifnum\draftcontrol=1 
  \def\@bibitem#1{%
    \st@bibitem{#1}\a@@label{#1}\ignorespaces} 
  \def\@lbibitem[#1]#2{%
    \st@lbibitem[#1]{#2}\a@@label{#2}\ignorespaces} 
  \def\a@@label#1{%
    \gdef\a@lab{\smash{\normalfont\small#1}} 
    \ifvmode 
      \if@inlabel 
        \global\setbox\@labels\hbox{%
          \llap{\a@lab\let\a@lab\relax 
                \kern\@totalleftmargin\kern\marginparsep}%
          \box\@labels}%
      \fi 
    \fi} 
\fi 

\documentclass[12pt,letterpaper]{article} 

\usepackage{amsmath,bm,amsfonts,amssymb,array,calc,amsthm,rotating}
\usepackage{epsfig,psfrag} 
\usepackage{graphicx}
\usepackage{color}
\usepackage[colorlinks=false]{hyperref}

\renewcommand\baselinestretch{1.25} 
\renewcommand\section{\@startsection {section}{1}{\z@}%
                                   {-3.5ex \@plus -1ex \@minus -.2ex}%
                                   {2.3ex \@plus.2ex}%
                                   {\normalfont\large\bfseries}} 
\renewcommand\subsection{\@startsection{subsection}{2}{\z@}%
                                   {-3.25ex\@plus -1ex \@minus -.2ex}%
                                   {1.5ex \@plus .2ex}%
                                   {\normalfont\normalsize\bfseries}} 
\renewcommand\subsubsection{\@startsection{subsubsection}{3}{\z@}%
                                   {-3.25ex\@plus -1ex \@minus -.2ex}%
                                   {1.5ex \@plus .2ex}%
                                   {\normalfont\normalsize\it}} 
\renewcommand\paragraph{\@startsection{paragraph}{4}{\z@}%
                                   {-3.25ex\@plus -1ex \@minus -.2ex}%
                                   {1.5ex \@plus .2ex}%
                                   {\normalfont\normalsize\bf}} 
\renewcommand\subparagraph{\@startsection{subparagraph}{5}{\z@}%
                                   {-1.25ex\@plus -1ex \@minus -.2ex}%
                                   {0ex \@plus .2ex}%
                                   {\normalfont\normalsize\it}}

\numberwithin{equation}{section}

\long\def\@makecaption#1#2{%
  \vskip\abovecaptionskip
  \sbox\@tempboxa{{\bf #1:} #2}%
  \ifdim \wd\@tempboxa >\hsize
    {\small\bf #1:} {\small #2}\par
  \else
    \global \@minipagefalse
    \hb@xt@\hsize{\hfil\box\@tempboxa\hfil}%
  \fi
  \vskip\belowcaptionskip}

\renewcommand*\l@section[2]{%
  \ifnum \c@tocdepth >\z@
    \addpenalty\@secpenalty
    \addvspace{.5em \@plus\p@}%
    \setlength\@tempdima{1.5em}%
    \begingroup
      \parindent \z@ \rightskip \@pnumwidth
      \parfillskip -\@pnumwidth
      \leavevmode \bfseries
      \advance\leftskip\@tempdima
      \hskip -\leftskip
      #1\nobreak\hfil \nobreak\hb@xt@\@pnumwidth{\hss #2}\par
    \endgroup
  \fi}
\renewcommand*\l@subsection{\addvspace{.0em \@plus\p@}\@dottedtocline{2}{1.5em}{2.3em}}
\renewcommand*\l@subsubsection{\addvspace{-.2em \@plus\p@}\@dottedtocline{3}{3.8em}{3.2em}}

\def\hepth#1{\href{http://xxx.arxiv.org/abs/hep-th/#1}{{arXiv:hep-th/#1}}}

\def\math#1{\href{http://xxx.arxiv.org/abs/math/#1}{{arXiv:math/#1}}}

\def\alggeom#1{\href{http://xxx.arxiv.org/abs/alg-geom/#1}{{arXiv:alg-geom/#1}}}
\def\arxiv#1#2{\href{http://xxx.arxiv.org/abs/#1}{{arXiv:#1 [#2]}}}

\definecolor{refcol}{rgb}{0.2,0.2,0.8}
\definecolor{eqcol}{rgb}{.6,0,0}
\definecolor{purple}{cmyk}{0,1,0,0}

\gdef\@citecolor{refcol}
\gdef\@linkcolor{eqcol}
\def\colorlinkspurple{\gdef\@urlcolor{purple}}
\def\colorlinksblue{\gdef\@urlcolor{blue}}
\def\colorlinksred{\gdef\@urlcolor{red}}

\def\ie{{\it i.e.}} 
\def\eg{{\it e.g.}}

\def\revise#1       {\raisebox{-0em}{\rule{3pt}{1em}}%
                     \marginpar{\raisebox{.5em}{\vrule width3pt\ 
                     \vrule width0pt height 0pt depth0.5em 
                     \hbox to 0cm{\hspace{0cm}{%
                     \parbox[t]{4em}{\raggedright\footnotesize{#1}}}\hss}}}} 

\newcommand\fnxt[1] {\raisebox{.12em}{\rule{.35em}{.35em}}\mbox{\hspace{0.6em}}#1} 
\newcommand\nxt[1]  {\\\fnxt#1}

\def\call         {{\cal L}} 
 
\def\caln         {{\cal N}}

\def\calt         {{\cal T}}

\def\calw         {{\cal W}}

\def\complex      {{\mathbb C}} 
 
\def\projective   {{\mathbb P}} 
 
\def\reals        {{\mathbb R}} 
\def\zet          {{\mathbb Z}} 
\def\CP{\complex\projective}
\def\RP{\reals\projective}

\def\del          {\partial} 
 
\def\ee           {{\it e}} 
\def\ii           {{\it i}}

\newcommand\topa[2]{\genfrac{}{}{0pt}{2}{\scriptstyle #1}{\scriptstyle #2}}

\def\sqr#1#2{{\vcenter{\vbox{\hrule height.#2pt   
 \hbox{\vrule width.#2pt height#1pt \kern#1pt 
 \vrule width.#2pt}\hrule height.#2pt}}}}

\def\ph{\eta}

\def\Ipp{\mathord{\mathchar "0271 \kern-4.5pt \mathchar"0271}}

\def\res{\mathop{{\rm Res}}}
\def\Li{{\rm Li}}

\catcode`\@=12 

\begin{document} 


\title{Calculations for Mirror Symmetry with D-branes}

\pubnum{%
CERN-PH-TH/2009-056 \\
\arxiv{0904.4905}{hep-th}}
\date{April 2009}

\author{
Johannes Walcher \\[0.2cm]
\it \href{http://ph-dep-th.web.cern.ch/ph-dep-th/}{PH-TH Division, CERN} \\
\it Geneva, Switzerland}

\Abstract{
We study normal functions capturing D-brane superpotentials on several one- 
and two-parameter Calabi-Yau hypersurfaces and complete intersections in weighted 
projective space. We calculate in the B-model and interpret the results using 
mirror symmetry in the large volume regime, albeit without identifying 
the precise A-model geometry in all cases. We identify new classes of 
extensions of Picard-Fuchs equations, as well as a novel type of topology
changing phase transition involving quantum D-branes. A 4-d domain wall which 
is obtained in one region of closed string moduli space from wrapping a 
four-chain interpolating between two Lagrangian submanifolds is, for 
other values of the parameters, represented by a disk ending on a single 
Lagrangian. 
}

\makepapertitle

\body

\version\versionno

\vskip 1em

\tableofcontents
\newpage

\section{Introduction}

In this paper, we study the contribution of background D-branes to the spacetime
superpotential for closed strings in type II string compactifications on compact
Calabi-Yau threefolds. Our focus is the extension of the methods developed in
\cite{open,mowa} towards making contact with the standard set of multi-parameter
models studied in the context of closed string mirror symmetry, and first in
\cite{cdfkm,cfkm,klemmyau1,klemmyau2}.

The superpotential on the D-brane worldvolume is an interesting quantity
to study, from many points of view, and has applications in all areas of D-brane
physics and mathematics. Generally speaking, one expects a holomorphic functional 
$\calw(u;z)$ on the infinitesimal open string state space, coordinatized by $u$, 
with parametric dependence on closed string moduli $z$. The expansion around the
critical points of $\calw$ with respect to $u$ should govern the low-energy interactions 
in the corresponding $\caln=1$ string vacuum. This description comes with the 
important caveat that it is hard to know how to compute invariantly off-shell 
(or relatedly that the physical couplings depend on the K\"ahler potential), but 
is an essential tool towards understanding D-branes on Calabi-Yau manifolds. For 
a very short sampling of early literature on this subject, see 
\cite{wittencs,vafaextends,bdlr,kklm,calin}. 

Dualities shed light on some of these questions. In particular, the relation of 
open topological strings to Chern-Simons theory and M-theory gives an interpretation 
of perturbative string amplitudes (of which the superpotential is the tree-level data) 
in terms of knot invariants and counting of BPS states, respectively \cite{oova}. This 
strategy opened the way to a quantitative understanding of D-brane superpotentials 
on non-compact geometries \cite{agva}, subsequently leading to many spectacular 
developments in topological string theory, see for example \cite{topvertex}. 
It has also been understood how these superpotential computations 
fit into a special geometry formalism \cite{mayr,lmw}. On the other hand, these results 
were, at least initially, restricted to non-compact setups and it has not been clear 
throughout how the compact case would be covered. More evidence is desirable to
further stabilize the status of the superpotential as a numerical invariant of the
D-brane configuration space. This should also help to reconnect with the algebraic
and categorical approaches developed for instance in \cite{fooo,aska}. 

The main lesson of \cite{open,mowa} is that already by just restricting the superpotential 
to the critical points,
\begin{equation}
\eqlabel{main}
\calw|_{\del_u\calw=0}
\end{equation}
one obtains a rather non-trivial invariant attached to a general, including compact, 
D-brane configuration. The quantity \eqref{main}, which depends on discrete open
string, and continuous closed string moduli, has a classical mathematical meaning in 
the B-model \cite{mowa}, as well as an enumerative interpretation in the A-model \cite{psw}.
Following the quintic, a handful of examples have now been worked out \cite{krwa,knsc1}. 
Progress on the relation to the framework of \cite{lmw} has also been made, see
\cite{joso,ghkk,ahmm,newpaper}.

In the present paper, we will continue to work with the quantity \eqref{main}, and
give examples of some further properties that appear over more complicated 
(multi-dimensional) moduli spaces. We will also touch on issues of compactification
of moduli, monodromy, and the open extension of the mirror map.

The bulk of our study proceeds by the analysis of examples. We will however begin
in section \ref{basic} with recalling the basic setup for the computation of the 
Picard-Fuchs equations in complete intersection Calabi-Yau as well as their extension 
to the open string sector. As in \cite{mowa}, the D-brane configurations that we study are
captured by a collection of holomorphic curves that reside at the intersection of the
Calabi-Yau with certain hyperplanes. This discussion will be followed by our first 
new example, based on the intersection of two cubics in $\projective^5$. It has the feature 
that the curves themselves are not complete intersection, but is otherwise qualitatively 
very similar to \cite{open,mowa,krwa}. Moreover, the enumerative predictions have been 
checked in the A-model, giving further support to the entire framework.

We will stay with one-parameter complete intersections in section \ref{second}. It
has been noted that among the fourteen hypergeometric one-parameter Calabi-Yau Picard-Fuchs 
equations, ten admit an extension by the same algebraic inhomogeneity as in \cite{open} 
that is sensible in the sense that the extending solution has an integral 
Ooguri-Vafa expansion. Of those ten, four are geometrically realized by hypersurfaces 
in weighted projective space, and the relevant D-brane geometries were 
identified in A- and B-model in \cite{open,mowa,krwa,knsc1}. (The integrality can 
then be a formal consequence of \cite{scvo}.) We will here supply the B-model 
branes for most of the remaining cases. Moreover, we will find other algebraic 
extensions that characterize a different vacuum structure ($\zet_p$ discrete Wilson 
line with $p>2$ as opposed to $p=2$ as in \cite{open}). Accompanying the discussion of the
two-parameter model, we will also find an extension of the Picard-Fuchs equation in one 
of the models for which the extension of \cite{open} did not make sense, see section
\ref{restrict}. The solution of this last extension is not hypergeometric.

The most involved computations are undertaken for the two-parameter model known as
$\projective_{11226}[12]$, for which analysis we will draw on \cite{cdfkm}. We 
summarize our D-brane geometry in section \ref{twopar}, and show that we obtain an integral
instanton expansion around the appropriate large volume point. In section \ref{modulispace},
we discuss the structure of the moduli space. When studying D-branes using \eqref{main}, 
one expects that the combined open-closed moduli space is generically a multi-covering 
of the closed string moduli space branched over the discriminant
locus over which the (discrete) D-brane moduli space becomes singular. Perhaps the
most interesting feature is that this D-brane discriminant, being of codimension one,
generically intersects the compactification divisor of the underlying closed string
moduli space. This intersection need not be transverse and one expects that interesting
physics will take place at these new types of singularities. (Most of the mathematics
should be in place, although the singularities of normal functions over multi-parameter
moduli spaces remain an active field of research, see, \eg, \cite{mattrecent} for
a recent survey.) 

We will analyze in detail only one of these new structures, where the D-brane discriminant
enters the large volume region. The first step here is an additional blowup of 
the moduli space described in \cite{cdfkm}. The coordinate on the exceptional divisor 
corresponds to the quantum volume (BPS tension) of a domainwall that interpolates 
between certain vacua of our D-brane geometry. As we will see, there are then two regimes 
in which the open-closed string background admits a classical geometric 
interpretation in the A-model. In one of them, the domainwall is represented in the A-model
by a 4-chain interpolating between two Lagrangian submanifolds, and in the other, by a 
disk ending on a single Lagrangian. (In the B-model, the domainwalls are always 
represented by 3-chains suspended between holomorphic curves.) The smooth interpolation 
between the two regimes constitutes a new instance of a topology changing
transition, of the type first observed in \cite{agm,wittenphases}. Note that we are
able to make these assertions without having identified the actual D-brane
configuration in the A-model. We will however discuss which qualitative features
this geometry must have in order to be consistent with the B-model and mirror symmetry.
A linear sigma model description of the phenomenon would be desirable.

\section{Overview and a Simple Example}
\label{basic}

In this paper, we study Calabi-Yau geometries that are obtained from the intersection
of the zero locus of a collection of polynomials $(W_j)_{1\le j\le n-3}$ in variables
$(x_i)_{1\le i\le n+1}$. The $x_i$ are homogeneous coordinates on
weighted projective space $\projective_{w_1,\ldots,w_{n+1}}^n$, and the $W_j$
are assumed to be homogeneous of degree $d_j$ with respect to the scaling specified by
the $w_i$. This means that
\begin{equation}
v (W_j) = d_j W_j
\end{equation}
where $v$ is the Euler vector field
\begin{equation}
\eqlabel{euler}
v = \sum_{i} w_i x_i \frac{\del}{\del x_i}
\end{equation}
The complete intersection of the $\{W_j=0\}$ is Calabi-Yau if $\sum_{i=1}^{n+1} w_i = 
\sum_{j=1}^{n-3} d_j$.

As usual, the A-model geometry, which we will denote by $X$, is obtained by choosing 
the $W_j$ generic transversal, and appropriately resolving the loci where their
intersection meets the singularities of the weighted projective space. The 
A-model then depends on $h_{11}(X)\ge 1$ independent K\"ahler classes and, at
closed string tree level, captures the classical intersection ring $H^{\rm even}(X)$ 
together with its quantum corrections due to worldsheet instantons. The B-model geometry, 
consequently denoted by $Y$, has several equivalent descriptions. We will use 
the version going back to Greene-Plesser \cite{grpl} in which $Y$ is the resolution of 
the quotient of a particular family (of dimension $h_{12}(Y)$) of manifolds $\cap\{W_j=0\}$ 
by a certain maximal discrete group of phase symmetries preserving the Calabi-Yau 
condition. The observables in the B-model originate mathematically from the variation 
of Hodge structure associated with the family. Mirror symmetry identifies A- and B-model 
and, in particular, $h_{11}(X)=h_{12}(Y)$.

We will study in practice only cases with $h_{11}(X)=h_{12}(Y)=1$ or $2$.
Just as the results of \cite{mowa,krwa}, the structure visible in the examples of the 
present paper is compatible with an application of methods of toric geometry 
\cite{batyrev} to compact Calabi-Yau geometries with D-branes. This was 
recently studied in \cite{ahmm}.

\subsection{Picard-Fuchs equations}

The moduli spaces governing closed string mirror symmetry can be studied by
computing the periods of the holomorphic three-form on $Y$, \ie
\begin{equation}
\eqlabel{periods}
\varpi(z) = \int_{\Gamma} \Omega
\end{equation}
where $\Omega\in H^{3,0}(Y)$, and $\Gamma\in H_3(Y;\zet)$. We have here summarily 
denoted the complex structure moduli of $Y$ by $z$.
They appear in certain combinations as parameters in the defining polynomials $W_j$. 
A good deal of information about the periods \eqref{periods} can be obtained from 
the differential equations that they satisfy as functions of the $z$. These differential 
equations originate from the fact that taking derivatives of $\Omega$ with respect to 
the parameters generates other elements of the third cohomology $H^3(Y)$. The latter 
being finite-dimensional results in cohomological relations amongst the derivatives 
of $\varpi(z)$, known as the Picard-Fuchs differential ideal. For such considerations 
to make sense, it is important that the three-cycles $\Gamma$ against which we 
integrate the three-forms be topological, \ie, they can be chosen independent 
of the complex structure parameters $z$.

The periods then satisfy the Picard-Fuchs differential equations, but so does
any ($z$-independent) complex linear combination. Singling out an integral basis
requires additional information that can be obtained in part from considerations 
of monodromy, in particular at points of maximal unipotent monodromy, as well
as by comparison with explicit integration around carefully chosen cycles.

A useful algorithm to derive these Picard-Fuchs equations is the Griffiths-Dwork 
reduction method \cite{griffiths1}. For the complete intersections in weighted 
projective space as described above, we may represent the holomorphic three-form 
as a residue
\begin{equation}
\eqlabel{omega}
\Omega = \frac{|G|}{(2\pi\ii)^{3}} {\res}_{W_j=0} \frac{\omega}{\prod_j W_j}
\end{equation}
where $\omega$ is the $n$-form
\begin{equation}
\omega= \alpha(v) = \sum_i (-1)^{i-1} w_i x_i dx_1\wedge\cdots\widehat{dx_i}\cdots
\wedge dx_{n+1}
\end{equation}
obtained by contracting the ``virtual'' $n+1$-form 
\begin{equation}
\alpha = dx_1\wedge\cdots\wedge dx_{n+1}
\end{equation}
with the Euler vector field $v$ from eq.\ \eqref{euler}. In \eqref{omega}, we have
inserted the order of the discrete group, $G$, that relates the $\{W_j=0\}$ to $Y$,
as in \eg, \cite{quevedo}. To make the meaning of the residue in \eqref{omega} more 
explicit, given a three-cycle $\Gamma\subset \cap_j\{W_j=0\}$, we construct a 
``tube-over-cycle'', $T(\Gamma)$, by fibering an $n-3$-dimensional torus over 
$\Gamma$ that surrounds all $\{W_j =0\}$ sufficiently closely. Then,
\begin{equation}
\eqlabel{closely}
\int_\Gamma\Omega = \frac{|G|}{(2\pi\ii)^{n}}\int_{T(\Gamma)}
\frac{\omega}{\prod_j W_i}
\end{equation}
Periods of derivatives of $\Omega$ with respect to the parameters take very similar forms. 
The fundamental relation that allows the Griffiths-Dwork reduction is the identity between 
meromorphic 
forms in the ambient (weighted) projective space,
\begin{equation}
\eqlabel{fundamental}
d \Bigl(\frac{A^i\omega_i}{P}\Bigr) = 
\frac{\del_i A^i \omega}{P} - \frac{A^i\del_i P \omega}{P^2}
\end{equation}
where $P$ is any homogeneous polynomial of degree say $D$, the $A^i$ are polynomials of 
degree $D-\sum_{j\neq i} w_j$, and $\omega_i$ is the contraction
\begin{equation}
\omega_i = \omega(\del_i) = \alpha(v,\del_i)
\end{equation}
By analyzing the relations in the polynomial ideal $\complex[x_i]/\del W_j$, helped by
exploiting the discrete group action, the Griffiths-Dwork method delivers differential 
operators $\call_{GD}$ together with meromorphic $n-1$-forms $\tilde\beta$ such that
\begin{equation}
\eqlabel{suchthat}
\call_{GD}\Bigl(\frac{\omega}{\prod_j W_j}\Bigr) = d\tilde\beta
\end{equation}
It is important for us to keep in mind that the explicit form of the $\tilde\beta$ depends 
on a choice of representatives for cohomology and relations.

Of course, for complete intersections of the type described above, the Picard-Fuchs 
equations can be obtained more efficiently by an appropriate extension of the 
GKZ differential system associated with the ambient toric variety \cite{bavs,klemmyau1,klemmyau2}. 
The equations are much simpler to solve in the resulting form that exposes the hypergeometric
structure of the solutions, so it is convenient to rewrite the equations in this fashion. 
In the examples, we will chose bases of relations such that the Griffiths-Dwork, $\call_{\rm GD}$, 
and hypergeometric, $\call$, operators are simply related by (possibly $z$-dependent) 
normalization factors $\caln_{\rm GD}$ and $\caln$.
\begin{equation}
\eqlabel{conjugate}
\call \caln = \caln_{\rm GD} \call_{\rm GD} 
\end{equation}
The normalization $\caln_{\rm GD}$ is of course irrelevant for computations
of ordinary periods, but becomes essential in the context of open string
computations, to which we now turn.

\subsection{Extensions}

The basic idea behind our computations is to study the open string observable 
\eqref{main} via its representation as a chain integral
\begin{equation}
\eqlabel{nonsense}
\calw_C = \int^C \Omega
\end{equation}
Here, $C\subset Y$ is a holomorphic curve representing the corresponding critical point
of $\calw$, and it is understood that to really carry out the integral, we should
choose a pair of homologically equivalent curves, and integrate $\Omega$ over
a bounding three-chain. This corresponds in \eqref{main} to the computation of BPS 
domainwall tensions as differences of superpotentials between critical points. 
Mathematically, such objects are known as normal functions \cite{griffiths1}. We 
refer to \cite{mowa} for a discussion of the applicability of \eqref{nonsense}. 

The chain integral \eqref{nonsense} is studied via its own differential equations,
which is an inhomogeneous extension of the ordinary Picard-Fuchs system. As we will
review, the inhomogeneous term results from a local computation around the curve $C$,
and does not depend on where we begin the integral. If one is interested in studying
domainwalls, and in particular for global consistency over the moduli space, one
also needs to fix the solution of the homogeneous equation, that can be freely added
to $\calw_C$, up to integral periods. As for the periods, this requires additional
information such as appropriate boundary conditions.

In the paper \cite{mowa}, an interesting D-brane configuration was obtained via 
some detours as a certain matrix factorization of the (single) polynomial defining 
a quintic hypersurface. The holomorphic curves needed for the computation of 
\eqref{nonsense} were representatives of the second algebraic Chern classes of these 
matrix factorizations. A similar strategy was pursued in \cite{krwa,knsc1} for
the other one-parameter hypersurfaces. The selection principle for the matrix
factorizations in \cite{mowa,krwa} was a conjectural mirror relation to the
Lagrangian submanifolds given as real slices of the corresponding A-model geometry. 
(In \cite{knsc1}, a different scheme was used, see \cite{knsc2} for a possible
extension to multi-parameter models.) 

In our examples, we will instead directly specify the curves. We do this first of all 
because the matrix factorization description of B-branes for complete intersections is 
more complicated \cite{hhp}, and second of all, because we do not have a confident conjecture 
about relevant mirror pairs of D-brane configurations. This is due in part to the absence 
of a Gepner point in the moduli space of these examples where mirror symmetry could have 
been based on an exactly solvable conformal field theory.

What we will borrow from \cite{mowa} is that the computation of the inhomogeneous
term is possible when the curves are components of the intersection of the $\{ W_j=0\}$ 
with two hyperplanes. (By hyperplane in a weighted projective space, we mean a subspace
linear in at least one coordinate.) Thus, in each of our examples, we will choose two 
such hyperplanes, $P_1$ and $P_2$, such that the intersection
\begin{equation}
\eqlabel{intersection}
\bigl(\cap_j \{W_j=0\}\bigr) \cap P_1\cap P_2 = \cup_{i} C_i
\end{equation}
decomposes into several component curves $C_i$. (If there were only one component,
the relevant integrals would all vanish automatically.) We will make the construction
such that the hyperplanes and the curves deform smoothly as we vary the
complex structure parameters $z$. (More precisely, we allow the possibility of
degeneration at co-dimension one discriminant loci, see below.) 

The hyperplanes $P_1$ and $P_2$ are typically not invariant under the action of the discrete
group $G$, and the curves on $Y$ really come from the orbits under that action. However,
the stabilizer can be non-trivial, and the curves will then intersect the singularities
that are resolved in the construction of $Y$. In \cite{mowa}, this was analyzed
carefully on the mirror quintic, and it was shown that the net effect is to divide
the final result by the order of the stabilizer, $S\subset G$. This amounts to replacing 
$|G|$ in \eqref{omega} by the length of the orbit, $|O|=|G|/|S|$, to which the respective curve 
belongs. This is the prescription that we shall assume.

In such a setup, there are in principle two ways to obtain a non-trivial normal
function, as observed in \cite{krwa}. We can compare via the chain integral \eqref{nonsense} 
either two component curves in the intersection \eqref{intersection}, averaged over
$G$, or the same components in two different $G$-orbits. In either case, the problem
at hand is the computation of the inhomogeneous term that results from the application
of a Picard-Fuchs operator to the chain integral \eqref{nonsense}.

So let us finally explain how we compute this inhomogeneous term. We exploit the
fact that while $\tilde\beta$ in \eqref{suchthat} is a meromorphic $n-1$-form, the 
curves in \eqref{intersection} are contained in $n-2$-dimensional linear subspaces of
the ambient projective space. Thus, by laying the tube around the curves inside of 
$P_1\cap P_2$ as much as possible, the computation of
\begin{equation}
\int_{T(C_i)} \tilde\beta
\end{equation} localizes to the points $\{p_1,p_2,\ldots\}$ where $C_i$ intersects
one of the other components. In a local neighborhood $U_k$ of each $p_k$, we chose 
a local parameterization of $C_i$, and $n-3$ normal vectors $n_j$ that point inside 
of $P_1\cap P_2$ outside of $U_k$. To guarantee that we are surrounding all $\{W_j=0\}$, 
it is most convenient to arrange the $n_j$ such that on $U_k$,
\begin{equation}
\begin{split}
n_j (W_j) &> 0\,, \\
n_i(W_j) &=0 \,,\qquad \text{for $i\neq j$}
\end{split}
\end{equation}
By perturbing the curve in the direction $\sum\epsilon_j n_j$, with each $\epsilon_j$
encircling the origin in the complex plane, the integral of $\tilde\beta$ around $C_i$ 
is computed from a combination of $n-2$ angular integrals and one radial integral, 
in a simple generalization of \cite{mowa}. Note that in this residue computation, 
the angular integrals combine with the $(2\pi\ii)^{-n}$ in \eqref{closely} to leave 
us with an overall $1/(2\pi\ii)^2$ characteristic of a normal function associated 
with a curve on a Calabi-Yau threefold.

We have thus computed
\begin{equation}
\eqlabel{first}
\frac{|O|}{(2\pi\ii)^n}\int_{T(C_i)} \tilde \beta
\end{equation}
as a first contribution to the inhomogeneous term in the Picard-Fuchs equation.
However, as anticipated, there is in general a second contribution, which 
originates from the action of the differential operator on the three-chain.
(In distinction to three-cycles, the three-chains are not annihilated by the
Gauss-Manin connection. Note that there is no invariant separation between the
two types of contributions.) As before, this contribution localizes to the curves,
and further to the intersection points $p_k$. To show how this is done in
practice, we let $n_z$ be a normal vector representing the first order variation 
of $C_i$ with respect to the complex structure parameter $z$. We then have, for 
example,
\begin{equation}
\eqlabel{intermediate}
\frac{\del^k}{\del z^k} \Bigl(\int^{C_i}\Omega\Bigr)
= \sum_{l=1}^{k}\frac{\del^{k-l}}{\del z^{k-l}}\Bigl(\int_{C_i} \bigl(\del_{z}^{l-1}
\Omega\bigr)(n_z)\Bigr) + \int^{C_i} \del_z^{k} \Omega
\end{equation}
The final term enters the Griffiths-Dwork reduction process, while the intermediate
integrals can be computed as residues as described above, and then safely 
differentiated. 

The total inhomogeneity is obtained by collecting \eqref{first} and terms of the 
form \eqref{intermediate}. As stressed above, to actually solve the inhomogeneous 
Picard-Fuchs equation, it is convenient to revert to its hypergeometric form via 
\eqref{conjugate}. 

\subsection{\texorpdfstring{Example. $\projective^5[3,3]$}{Example. P[3,3]}}

We begin our collection of examples with a model that is qualitatively similar
to the quintic studied in \cite{open,psw,mowa}. In particular, if we conjecture that
the D-brane configurations that we will specify below are mirror to the real
slices of the A-model, we obtain enumerative predictions that can be (and in fact
have been) checked by independent localization computations in the A-model.
The underlying manifold in the A-model is the intersection of two cubics in $\CP^5$,
which, in the notation of the previous subsections, corresponds to $n=5$, 
$w_1=\ldots=w_6=1$, and $d_1=d_2=3$. As first observed in \cite{libtei}, the 
mirror manifold can be represented by the quotient of a one-parameter family 
of bicubics,
\begin{equation}
\eqlabel{bicubdef}
Y = \{ W_1=0\,, W_2=0\} / G
\end{equation}
where $W_1$ and $W_2$ are the particular cubic polynomials
\begin{equation}
\begin{split}
W_1&= \frac{x_1^3}3 + \frac{x_2^3}3+ \frac{x_3^3}{3} - \psi x_4x_5x_6 \\
W_2&=\frac{x_4^3}3+\frac{x_5^3}3 + \frac{x_6^3}3 - \psi x_1x_2x_3 
\end{split}
\end{equation}
with the complex structure parameter $\psi$, and $G\cong \zet_3^2\times 
\zet_9$ is the maximal discrete group preserving $W_1$ and $W_2$, as well as the 
holomorphic three-form \eqref{omega}. By applying Griffiths-Dwork reduction as 
reviewed above, we obtain the Picard-Fuchs operator \cite{libtei}
\begin{equation}
\call_{\rm GD} = \frac{\psi^6-1}{64}\del_\psi^4 
+ \frac{7 \psi^6- 1}{32 \psi}\del_\psi^3 + \frac{55 \psi^6+1}{64\psi^2}\del_\psi^2
+\frac{65 \psi^6-1}{64 \psi^3}\del_\psi + \frac{\psi^2}{4}
\end{equation}
with an inhomogeneous term that in our scheme begins as
\begin{equation}
\begin{split}
\tilde\beta  = 
&-\frac{\psi^2 x_3\omega_3}{4W_1 W_2} - 
\frac{x_1^2 x_2x_3\omega_1}{192 \psi^3 W_1^2 W_2} + 
\frac{17 \psi^3 x_1^2x_2x_3\omega_1}{192 W_1^2 W_2}  
 - \frac{x_1x_2^2x_3\omega_2}{192\psi^3 W_1^2 W_2} \\& +
\frac{17 \psi^3 x_1x_2^2x_3\omega_2}{192 W_1^2 W_2}-
\frac{\psi^3x_1x_2x_3^2 \omega_3}{4 W_1 W_2^2} 
 -\frac{x_1x_2x_3^2\omega_3}{192 \psi^3 W_1^2 W_2} +
\frac{17 x_1x_2x_3^2\omega_3}{192 W_1^2 W_2} +
\cdots
\end{split}
\end{equation}
and contains a total of 64 similar such terms. By changing the normalization via 
\eqref{conjugate} with $\caln= \psi^2 $, $\caln_{\rm GD}=4/81$, 
we obtain the hypergeometric differential operator
\begin{equation}
\call = \theta^4 - 9 z (3\theta+1)^2(3\theta+2)^2
\end{equation}
where $z= (3\psi)^{-6}$, and $\theta=z \frac{d}{dz}$.

Turning to the specification of D-brane configurations, we consider the hyperplanes
\begin{equation}
\eqlabel{bicubhyper}
P_1 = \{x_1+x_2=0\} \,,\qquad
P_2 = \{ x_4+x_5=0\}
\end{equation}
The intersection of $P_1$, $P_2$ with $\{W_1=0, W_2=0\}$ is reducible. It contains the
line
\begin{equation}
C_0 = \{ x_1+x_2=0\,, x_4+x_5=0\,, x_3=0\,, x_6=0\}
\end{equation}
as well as two degree 4 curves, $C_+$ and $C_-$. Their homogeneous
ideal is generated by
\begin{equation}
\langle x_1+x_2, x_4+x_5, 
x_3^3 +3\psi x_4^2 x_6,
x_6^3+3\psi x_1^2 x_3,
x_3^4\pm 9 \psi^2 x_4^3 x_1, 
x_6^4\pm 9\psi^2 x_1^3 x_4 \rangle
\end{equation} 
Notice that these are not complete intersection curves. [The simplest way to understand
the curves is via their rational parameterization
\begin{equation}
\begin{split}
x_1 &= u^4\,,\quad x_2=-u^4\,, \quad x_3 = \alpha_1\sqrt{3\psi} u v^3 \\
x_4&= v^4 \,,\quad x_5 = -v^4 \,, \quad x_6 = \alpha_2 \sqrt{3\psi} u^3 v 
\end{split}
\end{equation}
where $(u,v)$ are homogeneous coordinates on $\projective^1$, and $\alpha_1, \alpha_2$ are
fourth roots of $\mp 1$ satisfying $\alpha_1^3+\alpha_2=\alpha_2^3+\alpha_1=0$.]
Implementing the residue computation sketched above (see \cite{mowa,krwa} for more 
details), we find
\begin{equation}
\int_{T(C_\pm)} \tilde \beta = \pm (2\pi\ii)^3 \frac{3}{32\psi^3}
\end{equation}
The action of the derivatives on the curves \eqref{intermediate} does not contribute
for our choice of tube and $\tilde\beta$. Finally, we need to collect the various
normalization factors. It is not hard to see that the discrete group acts by relating
$P_1$, $P_2$ to $9$ similar pairs of hyperplanes. The stabilizer of $C_{\pm}$ is
$\zet_9$. This leads to the inhomogeneous Picard-Fuchs equation in hypergeometric form
\begin{equation}
\eqlabel{inhombi}
\call \int^{C_\pm} \Omega = \frac{\pm1}{(2\pi\ii)^2} \frac{9}{8} \sqrt{z}
\end{equation}
We can solve this equation around $z=0$ by recalling the hypergeometric generating function,
\begin{equation}
\varpi(z;H) = \sum_{n=0}^\infty \frac{\Gamma(1+3(n+H))^2}{\Gamma(1+n+H)^6} z^{n+H}
\end{equation}
Indeed, we have, as a power-series in $z$, 
\begin{equation}
\call \varpi(z;H) = H^4 \frac{\Gamma(1+3H)^2}{\Gamma(1+H)^6} z^H
\end{equation}
Since
\begin{equation}
\call \del_H \varpi(z;H) = \del_H \call \varpi(z;H)\,,
\end{equation}
we find that the solutions of the homogeneous equation $\call\varpi(z)=0$
are the derivatives with respect to $H$,
\begin{equation}
\varpi_k(z) =  \del_H^k \varpi(z;0) \qquad \text{for $k=0,1,2,3$}
\end{equation}
A solution of the inhomogeneous equation \eqref{inhombi} is given by
\begin{equation}
\eqlabel{supocont}
\calw_{\pm}(z) = \pm\frac{1}{8} \varpi(z;1/2)
\end{equation}
as can be seen from the identity
\begin{equation}
\frac{\Gamma(1+3/2)^2}{\Gamma(1+1/2)^6} = \frac{36}{\pi^2}
\end{equation}
We should note that the actual domainwall tension might differ from
$\calw_+-\calw_-$ by a solution of the homogeneous equation. This could be studied by careful 
analytic continuation and computation of monodromies. The boundary condition at the 
point $\psi=0$ is somewhat more delicate to understand than on the quintic, since the
manifold \eqref{bicubdef} becomes singular there.

An A-model interpretation of the superpotential contribution \eqref{supocont} can be obtained
as usual by expanding the normalized expression
\begin{equation}
\hat\calw_\pm(q) = (2\pi\ii)^2 \frac{\calw_\pm(z(q))}{\varpi_0(z(q))}
\end{equation}
($\varpi_0$ is an integral period in the normalization \eqref{omega}, see \cite{quevedo}.)
in the appropriate flat coordinates
\begin{equation}
q = \ee^{2\pi\ii t} = \exp\Bigl(\frac{\varpi_1}{\varpi_0}(z)\Bigr)
\end{equation}
The first few terms are
\begin{equation}
\pm\hat\calw_\pm = 18 q^{1/2} + 182 q^{3/2} + \frac{787968}{25} q^{5/2} + \cdots 
\end{equation}
By the methods of \cite{psw}, it is known that the entire series $\hat\calw$ reproduces 
the open Gromov-Witten invariants counting maps from the disk to $(X,L)$, where $L$ 
is the real slice of the intersection $X$ of two generic cubics in $\projective^5$.
Notice that for $X=\{W_1=0, W_2=0\}$ at $\psi=0$, $L\cong \RP^3$, just as on the 
quintic.

Finally, we expand $\hat\calw$ using the multi-cover formula,
\begin{equation}
\eqlabel{defined}
\hat\calw_\pm = \pm\sum_{d,k\;{\rm odd}} \frac{N_d}{k^2} q^{k d/2}
\end{equation}
obtaining $N_1=18,N_3=180,N_5=31518, \ldots\,$.
According to the proposal of \cite{oova,open}, the $N_d$ count the degeneracy of BPS 
domainwalls separating the two 4-d $\caln=1$ vacua corresponding to the choice of
discrete Wilson line on $L\cong \RP^3$. Integrality of the $N_d$ as a mathematical 
theorem follows, as on the quintic, from the recent results of \cite{scvo}. Real
enumerative invariants in the sense of \cite{tadpole} are given by $N_d/2$.

\section{Some More One-parameter Models}
\label{second}

Before turning to the two-parameter model, we present here two more one-parameter
models with hypergeometric Picard-Fuchs equation. These are the simplest examples
for which we have found inhomogeneous terms different from the simple $\sqrt{z}$ 
extension prominent in all previously studied cases. We will be content with
solving the inhomogeneous Picard-Fuchs equations up to rational periods, and 
not work out the exact spectrum of domainwalls.

Closed string mirror symmetry for both models is discussed in detail in 
\cite{klth}. To abbreviate some of the formulas below, we introduce the hypergeometric
generating function
\begin{equation}
\varpi(z;H) = \sum_{n=0}^\infty \frac{\prod_{j=1}^{n-3} 
\Gamma(1+d_j(n+H))}{\prod_{i=1}^{n+1} \Gamma(1+w_i(n+H))}
z^{n+H}
\end{equation}
for given weights $w_i$ of the homogeneous coordinates, and degrees $d_j$ of the 
defining polynomials. It is also convenient to reserve a notation for the coefficient 
of the lowest order term of $\call\varpi(z;H)$
\begin{equation}
\eqlabel{xi}
\Xi(H) = H^4 \frac{\prod_{j=1}^{n-3} \Gamma(1+ d_j H)}
{\prod_{i=1}^{n+1}\Gamma(1+w_i H)}
\end{equation}

\subsection{\texorpdfstring{$\projective_{112112}[4,4]$}{P111122[4,4]}}

The B-model geometry is determined by
\begin{equation}
\eqlabel{112112def}
\begin{split}
W_1 &= \frac{x_1^4}4 + \frac{x_2^4}4+\frac{x_3^2}2 - \psi x_4 x_5x_6 \\
W_2 &= \frac{x_4^4}4+\frac{x_5^4}4+\frac{x_6^2}2- x_1x_2x_3
\end{split}
\end{equation}
and $G\cong \zet_2^2\times \zet_{16}$. The generators of $G$ can be taken to be,
in customary notation, \cite{klth}
\begin{equation}
\frac12(0,1,1,0,0,0)\,,\quad \frac12(0,0,0,0,1,1)\,,
\quad \frac{1}{16}(0,4,0,1,13,2)
\end{equation}
Griffiths-Dwork reduction with $\caln=\psi$, $\caln_{\rm GD}=1/64 \psi$ produces the 
Picard-Fuchs operator
\begin{equation}
\theta^4 - 16 z (4\theta+1)^2 (4\theta+3)^2
\end{equation}
where $z=(8\psi)^{-4}$.

\medskip\noindent
(i) We first exhibit the curves that produce the $\sqrt{z}$ extension familiar from the 
quintic. Consider the hyperplanes
\begin{equation}
\eqlabel{end}
P_1=\{x_1+\alpha_1 x_2\}\,\qquad  P_2= \{x_4+\alpha_2 x_5=0\}
\end{equation}
where $\alpha_1$, $\alpha_2$ are fourth roots of $-1$. These $16$ pairs of hyperplanes
decompose into two orbits of length $8$ under the action of $G$. The intersection of 
$W_1=0$, $W_2=0$ with $P_1$, $P_2$ splits into two components,
\begin{equation}
\begin{split}
C_+&= P_1\cap P_2\cap \{x_3=0,x_6=0\} \\
C_-&= P_1\cap P_2\cap \{x_3^3+8\psi^2\alpha_2^2\alpha_1 x_4^4 x_2^2=0,
x_2^2+2\psi\alpha_2 x_4^2 x_6=0,
x_6^2+2\alpha_1 x_2^2 x_3=0\}
\end{split}
\end{equation}
The residue computation delivers
\begin{equation}
\call_{\rm GD} \int^{C_\pm}\Omega =\pm \frac{|O|}{(2\pi\ii)^2} \frac{1}{2\psi}
\end{equation}
and gives the inhomogeneous Picard-Fuchs equation
\begin{equation}
\call \calw_\pm(z) = \pm \frac{4}{(2\pi\ii)^2}  \sqrt{z}
\end{equation}
As in the previous cases, this is solved by
\begin{equation}
\calw_\pm= \pm \frac {1}4\varpi(z;1/2)
\end{equation}
The normalization again comes out thanks to
\begin{equation}
\Xi(1/2) = \frac{1}{2^4} \,\frac{\Gamma(1+2)^2}{\Gamma(1+1/2)^4\Gamma(1+1)^2} = \frac{4}{\pi^2}
\end{equation}
The low-degree BPS invariants computed as above \eqref{defined} are $N_1=64$, $N_3=5568$, 
$N_5=4668864,\ldots\,$.
It is possible that these can be interpreted as real enumerative invariants as for 
the previously studied models \cite{open,krwa,knsc1}. Note also that since we have 
two $G$-orbits in the set of planes described above, the collection of curves shares
a second domainwall that is a rational linear combination of closed string periods.
This is similar to an observation on the degree $8$ hypersurface in \cite{krwa}.

\medskip
\noindent
(ii) Now consider the hyperplanes
\begin{equation}
P_1 = \{ x_1^2 +\alpha_1 \sqrt{2} x_3\}\,,\qquad
P_2 = \{ x_4^2+\alpha_2 \sqrt{2} x_6\}
\end{equation}
where $\alpha_1,\alpha_2=\pm \ii$. All pairs are related by the action of $G$,
so there is only one orbit, of length $4$. The intersection with $W_1=0$, $W_2=0$ 
splits into 4 components,
\begin{equation}
\begin{split}
C_{0} &= P_1\cap P_2\cap \{ x_2=0, x_5=0\}\\
C_\ph &= P_1\cap P_2\cap \{ W_1=0,W_2=0, x_2^5 = \ph
\frac{2^{5/2}\psi^{4/3}}{(-\alpha_2^{4}\alpha_1)^{1/3}} x_4^4 x_1 \} 
\end{split}
\end{equation}
where $\ph$ is a third root of unity keeping track of the root of
$\psi$ we are taking. This pattern of curve is signaling a $\zet_3$-symmetric 
collection of brane vacua. Indeed, the residue computation gives
\begin{equation}
\call_{\rm GD} \int^{C_\ph} \Omega = \frac{|O|}{(2\pi\ii)^2}
\Bigl( \frac{8}{27\ph^2\psi^{1/3}} + \frac{50}{27\ph\psi^{5/3}}\Bigr)
\end{equation}
[We also find
\begin{equation}
\call_{\rm GD} \int^{C_0} \Omega=0\,,
\end{equation} 
as necessary.]
Transformation to hypergeometric form yields
\begin{equation}
\call \int^{C_\ph} \Omega = \frac{1}{(2\pi\ii)^2}
\Bigl( \frac{8}{27}\ph z^{1/3}+\frac{800}{27} \ph^{2}z^{2/3}\Bigr)
\end{equation}
Noticing that
\begin{equation}
\Xi(1/3) = \frac{1}{3\pi^2}\,,
\qquad
\Xi(2/3) = \frac{100}{3\pi^2}
\end{equation}
(where $\Xi$ is defined in \eqref{xi}) we see that the solution can be expressed as
\begin{equation}
\calw_\ph = \frac{2}9 \bigl(\ph\varpi(z;1/3)+\ph^2\varpi(z;2/3)\bigr)
\end{equation}
We can now apply the standard mirror map to obtain the A-model expansion
\begin{equation}
\hat \calw_\ph = (2\pi\ii)^2\frac{\calw(z(q))}{\varpi_0(z(q))} =
24 \ph q^{1/3}+ 150 \ph^2 q^{2/3} + \frac{2571}2\ph^4 q^{4/3}
+\frac{417024}{25}\ph^5 q^{5/3}+ \cdots
\end{equation}
We see that the multi-cover formula in the present case
takes the form
\begin{equation}
\eqlabel{present}
\hat\calw_\ph = \sum_{3\nmid d,k} \frac{N_d}{k^2}\ph^{k d} q^{k d /3}
\end{equation}
with integral $N_d$ (as far as we have checked) counting domainwall degeneracies.
Understanding the precise geometric meaning of these invariants depends on identifying 
the A-model geometry mirror to our curves $C_\ph$. The symmetry suggests that
the corresponding Lagrangians $L$ have a factor $\zet_3\subset H_1(L)$ in their
first homology group, and $\ph$ is a discrete Wilson line. It seems unlikely (although 
it cannot be excluded) that these can be described as real slices of the complete 
intersection \eqref{112112def}. The first few non-trivial numbers are $N_1=24$, 
$N_2=144$, $N_4=1248$, $N_5=16680,\ldots$.

\subsection{\texorpdfstring{$\projective_{123123}[6,6]$}{P112233[6,6]}}

Here, the B-model geometry is determined by
\begin{equation}
\eqlabel{123123def}
\begin{split}
W_1 &= \frac{x_1^6}6 + \frac{x_2^3}3+\frac{x_3^2}2 - \psi x_4 x_5x_6  \\
W_2 &= \frac{x_4^6}6+\frac{x_5^3}3+\frac{x_6^2}2- x_1x_2x_3
\end{split}
\end{equation}
and $G\cong \zet_{36}$, whose generator we take to be
\begin{equation}
\eqlabel{Ggen}
g= \frac 1{36}(0,24,18,1,14,21)
\end{equation}
Griffiths-Dwork reduction with $\caln_{\rm hyper}=\psi$, $\caln_{\rm GD}=1/324 \psi^3 $
produces the Picard-Fuchs operator
\begin{equation}
\call= \theta^4 - 144 z (6\theta+1)^2(6\theta+5)^2
\end{equation}
with $z= 4^{-1}(6\psi)^{-6}$.

\medskip\noindent
(i) We here begin with the set of hyperplanes
\begin{equation}
\eqlabel{beginplanes}
P_1 =\{ x_1^2 + 2^{1/3} \alpha_1 x_2 =0\}
\,,\qquad
P_2 = \{ x_4^2 + 2^{1/3} \alpha_2 x_5 =0\}
\end{equation}
where $\alpha_1$, $\alpha_2$ are third roots of $-1$. The curves are
\begin{equation}
\begin{split}
C_0  &= P_1\cap P_2\cap \{ x_3=0,x_6=0\} \\
C_\ph &= P_1\cap P_2\cap\{ W_1=0,W_2=0,
x_3 = \ph \frac{2^{2/3} \psi^{2/3}}{(-\alpha_2^2\alpha_1)^{1/3}} x_4^2 x_1 \}
\end{split}
\end{equation}
where $\ph$ runs over third roots of unity. These curves have residue (up to an
overall phase that depends on the choice of hyperplane, see discussion in the
next paragraph)
\begin{equation}
\eqlabel{haveresidue}
\begin{split}
\call_{\rm GD} \int^{C_0}\Omega &= \frac{|O|}{(2\pi\ii)^2} 
\Bigl( - 2^{1/3} \psi-\frac{9}{2^{1/3} \psi}\Bigr)\\
\call_{\rm GD} \int^{C_\ph}\Omega &= \frac{|O|}{(2\pi\ii)^2}
\Bigl( \frac{2^{1/3}}{3} \psi+\frac{3}{2^{1/3} \psi}\Bigr)
\end{split}
\end{equation}
The group $G\cong\zet_{36}$ organizes the planes \eqref{beginplanes} into 
$3$ orbits of length $3$. In this example, $G$ also acts on the $\ph$-label 
in a non-trivial fashion. If $g$ is a generator as in \eqref{Ggen},
then $g^3$ acts within a given hyperplane by $\ph\to \ee^{2\pi\ii/3}\ph$. Note 
that this symmetry is consistent with the residues \eqref{haveresidue} being 
independent of $\ph$. Another consistency check on \eqref{haveresidue} is that
the sum of residues over all curves in a given hyperplane vanishes.
In the end, the $C_\ph$ come in $3$ orbits of length $9$, while the $C_0$
come in $3$ orbits of length $3$. This translates into
\begin{equation}
\call \int^{C_\ph}\Omega = -\call\int^{C_0}\Omega = \frac{1}{(2\pi\ii)^2}
\Bigl( \frac{2}{3}\tilde\ph z^{1/3} + 216\tilde\ph^2 z^{2/3}\Bigr)
\end{equation}
We have here reintroduced third roots of unity $\tilde\ph$ that keep track of 
the orbit of planes. In other words, $\tilde\ph$ depends on a combination of 
$\alpha_1, \alpha_2$ in \eqref{beginplanes}. In the present model,
\begin{equation}
\Xi(1/3) = \frac{3}{4\pi^2}
\,,
\qquad
\Xi(2/3) = \frac{243}{\pi^2}  \,.
\end{equation}
So a solution of the inhomogeneous Picard-Fuchs equation can be written as
\begin{equation}
\calw_{\tilde \ph} = 
\frac 2{9}\bigl(\tilde\ph \varpi(z;1/3)+\tilde\ph^2\varpi(z;2/3)\bigr)
\end{equation}
The A-model interpretation is the same as around \eqref{present}, with
$N_1=54$, $N_2=1080$, $N_4=216432$, $N_5=10094490, \ldots$.

\medskip\noindent
(ii)
Finally, we intersect with 
\begin{equation}
\eqlabel{finallyplanes}
P_1 = \{ x_1^3 + \alpha_1 \sqrt{3} x_3 =0\}\,,
\qquad
P_2 = \{ x_4^3 + \alpha_2\sqrt{3} x_6=0 \}
\end{equation}
where $\alpha_1,\alpha_2=\pm \ii$. There are now $5$ curves in each hyperplane,
\begin{equation}
\begin{split}
C_0 &= P_1\cap P_2\cap \{x_2=0, x_5=0\} \\
C_\ph &= P_1\cap P_2\cap \{W_1=0,W_2=0, x_2^2 = \ph 
\frac{\sqrt{3}\psi^{3/4}}{(\alpha_2^3\alpha_1)^{1/4}} x_4^3 x_1 \}
\end{split}
\end{equation}
where $\ph$ is a fourth root of unity.
\begin{equation}
\begin{split}
\call_{\rm GD}\int^{C_0}\Omega &= \frac{|O|}{(2\pi\ii)^2}\, 6 \\
\call_{\rm GD} \int^{C_\ph} \Omega &= \frac{|O|}{(2\pi\ii)^2} 
\Bigl( - \frac{9\sqrt{3}\ph^2 \psi^{3/2}}{32} - \frac 32  -\frac{147\sqrt{3}}{32\ph^2 \psi^{3/2}}
\Bigr)
\end{split}
\end{equation}
Here, $G$ organizes the hyperplanes \eqref{finallyplanes} into $2$ orbits of length $2$.
$g^2$ acts within a given plane by $\ph\to-\ph$. So the $C_\ph$ end up in $4$ orbits of length
$4$, and $C_0$ in $2$ orbits of length $2$. Thus,
\begin{equation}
\eqlabel{matches}
\begin{split}
\call \int^{C_0} \Omega &=  \frac{1}{(2\pi\ii)^2} \, 16 \sqrt{z} \\
\call \int^{C_\ph} \Omega &= \frac{1}{(2\pi\ii)^2}
\Bigl( \frac{1}{8} \ph^2 z^{1/4} - 8\sqrt{z} + 
882\ph^2 z^{3/4}\Bigr)
\end{split}
\end{equation}
The specialization of the hypergeometric coefficient \eqref{xi} here gives
\begin{equation}
\Xi(1/4)= \frac{1}{8\pi^2}
\,, \qquad
\Xi(1/2) = \frac{16}{\pi^2}
\,,\qquad
\Xi(3/4) =  \frac{882}{\pi^2}
\end{equation}
This matches the relative coefficients between $z^{1/4}$ and $z^{3/4}$ in \eqref{matches},
while the coefficient of $z^{1/2}$ is matched in a linear combination of the $C_0$
and $C_\ph$. To work out the entire spectrum of domainwall, it would be natural to
use the two linearly independent solutions
\begin{equation} 
\begin{split}
\calw_{\pm} &= \pm \frac{1}{4} \varpi(z;1/2) \\
\calw_{\tilde\ph} &= \frac{1}{4}\bigl( \tilde\ph\varpi(z;1/4) +
\tilde\ph^2 \varpi(z;1/2) + \tilde\ph^3\varpi(z;3/4)  \bigr)
\end{split}
\end{equation}
where $\tilde\ph$ is a fourth root of unity. The Ooguri-Vafa expansion for the two types
is, respectively
\begin{equation}
\begin{split}
\hat \calw_\pm &= \pm \sum_{2\nmid d,k} \frac{N_{2,d}}{k^2} q^{d k/2} \\
\hat \calw_{\tilde \ph} &= \sum_{4\nmid d,k} \frac{N_{4,d}}{k^2} \tilde\ph^{d k} q^{d k/4}
\end{split}
\end{equation}
The first few invariants are
\begin{equation}
\begin{split}
N_{2,1} &= 256\,,\;N_{2,3}=1742592\,,\; N_{2,5}= 65066366720\,,\ldots\\
N_{4,1} &= 32\,,\; N_{4,2}=248\,,\; N_{4,3}=2784\,,\; N_{4,5}=83680\,,\; N_{4,6}=1741896\,,\ldots 
\end{split}
\end{equation}

\section{A Two-parameter Model}
\label{twopar}

We now begin our investigation of D-brane superpotentials in the much-studied 
two-parameter model $\projective_{11226}[12]$. For the geometry of the closed
string moduli space, we rely on the treatment in \cite{cdfkm,klemmyau1}. 

\subsection{Data}

The generic degree $12$ hypersurface in $\projective_{11226}$ meets the singularities 
of the weighted projective space in a curve along which we have to blow up to 
produce the A-model geometry $X$. The resolution of singularities can be 
understood by giving the charges of the gauged linear sigma model fields
\begin{equation}
\eqlabel{glsm}
\begin{array}{c|cccccc|c}
& x_1 & x_2 & x_3 & x_4 & x_5 & x_6 & P \\\hline
h_1 & 0 & 0 & 1 & 1 & 3 & 1 & -6 \\
h_2 & 1 & 1 & 0 & 0 & 0 & -2 & 0 
\end{array}
\end{equation}
We will slightly depart from the notation of \cite{cdfkm}, and denote by $H_1$ 
the divisor class of $x_3=0$, $H_2$ the class of $x_1=0$. The exceptional
divisor is $E=H_1-2H_2$, and $h_1$, $h_2$ are the dual curve classes. We recall 
the classical intersection relations 
\begin{equation}
\eqlabel{intersect}
\begin{array}{c}
H_2^2 =0 \,, \qquad H_1^3 = 4\,,\qquad H_1^2H_2=2 \\
2h_1= H_1H_2\,, \qquad 2h_2 = H_1^2-2H_1H_2
\end{array}
\end{equation}
The mirror manifold is the two-parameter family of Calabi-Yau threefolds obtained
from the vanishing locus of the defining polynomial
\begin{equation}
\eqlabel{poly2par}
W = \frac{x_1^{12}}{12}+ \frac{x_2^{12}}{12} + \frac{x_3^6}6 + \frac{x_4^6}6 + \frac{x_5^2}2 - 
\psi x_1x_2x_3x_4x_5 - \frac{\phi}6 x_1^6 x_2^6
\end{equation}
after orbifolding with respect to the maximal group of phase symmetries 
$G=\zet_6\times\zet_6\times \zet_2$. One may work with the generators
\begin{equation}
\frac 16(0,5,1,0,0)\,,\qquad
\frac 16(0,5,0,1,0)\,,\qquad
\frac 12(0,1,0,0,1)
\end{equation}
The periods of the model are governed by a system of two Picard-Fuchs equations. 
The Griffiths-Dwork algorithm gives the following relations
\begin{equation}
\eqlabel{relations}
\begin{split}
\call_{{\rm GD},1} &= -\frac 16\del_\psi^3 + \psi^5\del_\psi^2\del_\phi
+\frac{1}{2\psi} \del_\psi^2 + 3 \psi^4 \del_\psi\del_\phi -
\frac{1}{2\psi^2} \del_\psi + \psi^3 \del_\phi \\
\call_{{\rm GD},2} &=  18(\phi^2-1) \del_\phi^2 +\frac{\psi^2}2\del_\psi^2
+ 6\psi\phi\del_\psi\del_\phi +\frac{3\psi}2\del_\psi +24 \phi
\del_\phi+\frac 12
\end{split}
\end{equation}
Together with appropriate three-forms $\tilde\beta_1,\tilde\beta_2$ whose explicit
form we shall suppress. The transformation to hypergeometric form is accomplished 
by conjugating the $\call_{{\rm GD},i}$ as in \eqref{conjugate} with 
$\caln_1=\caln_2=\psi$, and
\begin{equation}
\caln_{{\rm GD},1}= \frac{\phi}{36\psi^2} 
\,,\qquad\caln_{{\rm GD},2}= \frac{\psi}{72}
\end{equation}
One obtains
\begin{equation}
\eqlabel{hypertwo}
\begin{split}
\call_1 &= \theta_1^2(\theta_1-2\theta_2) - 8 z_1(6\theta_1+1)
(6\theta_1+3)(6\theta_1+5)  \\
\call_2 &= \theta_2^2 - z_2 (2\theta_2-\theta_1)(2\theta_2-\theta_1+1)
\end{split}
\end{equation}
with $z_1 = -3^{-3}2^{-6}\phi \psi^{-6}$, $z_2= (2\phi)^{-2}$, and $\theta_i=z_i\frac{d}{dz_i}$.
The solutions of \eqref{hypertwo} as power series around $z_1=z_2=0$ can be obtained 
from the hypergeometric generating function,
\begin{multline}
\eqlabel{gentwo}
\varpi(z_1,z_2;H_1,H_2) = \sum_{n_1,n_2=0}^\infty  z_1^{n_1+H_1} z_2^{n_2+H_2}\\
\frac{\Gamma(1+6(n_1+H_1))}{\Gamma(1+n_2+H_2)^2\Gamma(1+n_1+H_1)^2
\Gamma(1+3(n_1+H_1))\Gamma(1+n_1+H_1-2(n_2+H_2)))} 
\end{multline}
by differentiation,
\begin{equation}
\eqlabel{soltwo}
\begin{split}
\varpi_0 &= \varpi(z_1,z_2;0,0) \\
\varpi_{h_1} &= \del_{H_1} \varpi(z_1,z_2;0,0)\\
\varpi_{h_2} &= \del_{H_2} \varpi(z_1,z_2;0,0) \\
\varpi_{H_1} &= \bigl(2\del_{H_1}^2+2\del_{H_1}\del_{H_2}\bigr) \varpi(z_1,z_2;0,0) \\
\varpi_{H_2} &= \del^2_{H_1} \varpi(z_1,z_2;0,0) \\
\varpi_{X} &= -\Bigl(\frac 23\del_{H_1}^3+\del^2_{H_1}\del_{H_2}\Bigr) \varpi(z_1,z_2;0,0)
\end{split}
\end{equation}
Recall how these solutions reflect the GLSM charges \eqref{glsm} and the intersection 
relations \eqref{intersect}. The closed string mirror map around the large volume
point identifies the K\"ahler parameters as
\begin{equation}
\eqlabel{standardmir}
h_1= t_1  = \frac{1}{2\pi\ii} \frac{\varpi_{h_1}}{\varpi_0} \,,
\qquad
h_2 =t_2  = \frac{1}{2\pi\ii} \frac{\varpi_{h_2}}{\varpi_0}
\end{equation}

\subsection{Curves and residues}

As in the one-parameter examples studied in the previous sections, there are several 
possibilities for intersecting \eqref{poly2par} with two hyperplanes such that the 
resulting plane curve splits in a non-trivial way in several components, thus 
realizing the basic pattern of \cite{mowa}. Presently the most interesting 
curves are those obtained from the hyperplanes
\begin{equation}
\eqlabel{inthyper}
P_1 = \{ x_3 + 2^{-1/6} \alpha_1 x_1^2=0 \}  \,,
\qquad
P_2 = \{ x_4 + 2^{-1/6} \alpha_2 x_2^2=0 \} 
\end{equation}
where $\alpha_1$, $\alpha_2$ are sixth roots of $-1$. The intersection of $P_1\cap P_2$ 
with $\{W=0\}$ splits in two components,
\begin{equation}
\{ x_5 = \alpha_\pm x_1^3 x_2^3 \}
\end{equation}
where $\alpha_\pm$ are the two solutions of the quadratic equation
\begin{equation}
\eqlabel{fun}
\alpha_\pm^2 - 2^{2/3}\psi \alpha_1\alpha_2 \alpha_\pm - \frac{\phi}3 = 0
\end{equation}
In a way by now familiar, the discrete group $G$ permutes the planes in \eqref{inthyper}.
There are $3$ orbits of length $12$. We will label the resulting curves by
$C_{(\ph,\pm)}$, where $\ph$ is a third root of unity encoding the orbit of planes,
and $\pm$ refers to the choice of root in \eqref{fun}. 

Computation of the residues for the two relations in \eqref{relations} gives
\begin{equation}
\eqlabel{verysimilar}
\begin{split}
\call_{{\rm GD},1} \int^{C_{(\ph,\pm)}}\Omega &= 
-\frac{|O|}{(2\pi\ii)^2} 
\frac{2^{5/3}\ph \psi^3}{3(2\alpha_\pm-2^{2/3} \ph\psi)^3} 
\\
\call_{{\rm GD},2} \int^{C_{(\ph,\pm)}}\Omega &=
\frac{|O|}{(2\pi\ii)^2}
\frac{2^{5/3} \ph}{2\alpha_\pm-2^{2/3}\ph\psi}
\end{split}
\end{equation}
Collecting all the factors, and solving \eqref{fun}, we can transform to hypergeometric
form
\begin{equation}
\eqlabel{fornow}
\begin{split}
\call_1\int^{C_{(\ph,\pm)}}\Omega &= \pm \frac{1}{(2\pi\ii)^2} \frac{4\ph y}{6(1-4\ph y)^{3/2}} \\
\call_2\int^{C_{(\ph,\pm)}}\Omega &= \pm \frac{1}{(2\pi\ii)^2} \frac{1}{3(1-4\ph y)^{1/2}} 
\end{split}
\end{equation}
where we have introduced the variable
\begin{equation}
\eqlabel{crucialrole}
y = \Bigl(\frac{z_1}{z_2}\Bigr)^{1/3} = - \frac{2^{2/3}\phi}{12\psi^2}
\end{equation}
This combination will play a crucial role in the following discussion. Its precise
geometric role will be clarified during our discussion of the combined open-closed
moduli space in section \ref{modulispace}. For now, we proceed with solving 
\eqref{fornow}. 

\subsection{Solutions}

To get a good power series expansion, we transform to the independent variables 
$z_2,y$. We work with $\ph=1$, $\pm=+$, and absorb the factor of $(2\pi\ii)^2$ 
into $\calw\sim \int^C\Omega$. With $\theta_y=y\frac{d}{dy}$, we have
\begin{equation}
\eqlabel{goodpower}
\begin{split}
9 \call_1 \calw &= \bigl(\theta_y^2(\theta_y-2\theta_2) - 
72 y^3 z_2(2\theta_y+1)(2\theta_y+3)(2\theta_y+5)\bigr) \calw
= 3\,\frac{4 y}{2(1-4y)^{3/2}}
\\
9 \call_2 \calw &= \bigl( (\theta_y-3\theta_2)^2 - 9 z_2(2\theta_2-\theta_y)
(2\theta_2-\theta_y+1)\bigr)\calw 
= 3\, \frac{1}{(1-4y)^{1/2}}
\end{split}
\end{equation}
This form of the equations gives us the opportunity to verify, as a consistency check 
on our computations so far, that the system of partial differential equations \eqref{fornow} 
is integrable. Indeed, restricted to $z_2=0$, the Picard-Fuchs operators satisfy the 
relation
\begin{equation}
\call_1 = \theta_y \call_2
\end{equation}
and clearly, the inhomogeneities are consistent with this relation. In fact, expanding
\begin{equation}
\eqlabel{expanding}
\frac{1}{\sqrt{1-4y}} = \sum_{m=0}^\infty \frac{\Gamma(1+2m)}{\Gamma(1+m)^2} y^m
\end{equation}
we can integrate straightforwardly to obtain the solution at $z_2=0$,
\begin{equation}
\eqlabel{z20sol}
\calw(y,0) = 3\Bigl[\frac 12 (\log y)^2 + \sum_{m=1}^\infty
\frac{\Gamma(1+2m)}{m^2\Gamma(1+m)^2} y^m\Bigr]
\end{equation}
Not surprisingly, this series can be rewritten using the dilogarithm function. But let us 
put off a discussion of its analytic properties as a function of $y$ until section 
\ref{phasetransition}.

Using \eqref{z20sol} as a first step, one may find a representation of the higher order 
terms by solving the appropriate recursion relations. With the ansatz
\footnote{Summation indices will always be assumed to run over non-negative
integers, with further restrictions as indicated.}
\begin{equation}
\eqlabel{someof}
\begin{split}
\calw(y,z_2) =& 3\Bigl[ \frac 12 (\log y)^2 \sum_{\topa{m\in 3\zet,n}{2n\le m\le 3n}} 
a_{m,n} y^m z_2^n 
 + \log y \sum_{\topa{m\in 3\zet,n}{m\le 3n}} b_{m,n} y^m z_2^n \\
&+ \sum_{m\in 3\zet,n} c_{m,n} y^m z_2^n 
+ \sum_{m\notin 3\zet,n} d_{m,n} y^m z_2^n \Bigr]
\end{split}
\end{equation}
one obtains
\begin{equation}
\begin{split}
a_{m,n} &= \frac{\Gamma(1+2m)}{\Gamma(1+\frac m3)^2\Gamma(1+m)\Gamma(1-\frac m3+n)^2
\Gamma(1+m-2n)} \\
b_{m,n} &=a_{m,n}\bigl[{\textstyle 2\Psi(1+2m) - \frac 23\Psi(1+\frac m3)-\Psi(1+m)+
\frac 23\Psi(1-\frac m3+n) -\Psi(1+m-2n)}\bigr] \\
c_{m,n} &=\frac 12 a_{m,n}\Bigl[{\Bigl(\frac{b_{m,n}}{a_{m,n}}\Bigr)^2} + \\ 
&\quad 
{\textstyle 4\Psi'(1+2m) - \frac 29\Psi'(1+\frac m3)-\Psi'(1+m)-\frac 2 9\Psi'(1-\frac m3+n)
-\Psi'(1+m-2n)} \Bigr] \\
d_{m,n} &= \frac{\Gamma(1+2m)\Gamma(\frac m3-n)^2}
{9 \Gamma(1+\frac m3)^2\Gamma(1+m) \Gamma(1+m-2n)}
\end{split}
\end{equation}
where $\Psi$ is the digamma function, $\Psi'$ its derivative, and it is understood that
the formulas for $b_{m,n}$ and $c_{m,n}$ require a certain limit when the arguments of
the $\Gamma$-functions hit the poles. In particular, some of the restrictions on the
summation indices in \eqref{someof} are automatic. After noting that
\begin{equation}
\eqlabel{rationalize}
c_{0,0} = \frac 12\Psi'(1) (2-\frac 49) = \frac{7\pi^2}{54}
\end{equation}
we may rationalize the $c_{m,n}$ by subtracting the appropriate multiple
of $\varpi_0$ from $\calw$. It is in fact natural to rewrite this solution in a 
more suggestive fashion. The lift of \eqref{glsm} appropriate for our new variables 
$y,z_2$ is the table
\begin{equation}
\begin{array}{c|cccccc|c}
h& -\frac13 & -\frac13 & \frac 13 & \frac 13 & 1 & 1 & -2 \\
l& 1 & 1 & 0 & 0 & 0 & -2 & 0
\end{array}
\end{equation}
[This transformation is related to the following (non-integral) change of basis of cohomology
\begin{equation}
\eqlabel{curverels}
\begin{array}{c}
H=3 H_1 \,,\qquad L= H_1+H_2 \\[.1cm]
\displaystyle h = \frac 13 (h_1-h_2)=\frac{HL}6-\frac{2H^2}{27} \,,\qquad l=h_2=\frac{H^2}6 -
\frac{HL}3\\[.3cm]
\displaystyle H_1 =\frac{H}3 \,,\qquad H_2=L-\frac H3 \\[.1cm]
h_1=3h+l\,,\qquad h_2=l ]
\end{array}
\end{equation}
The generating function of solutions now takes the form
\begin{multline}
\eqlabel{withwhich}
\tilde\varpi(y,z_2;H,L) = \sum_{m\in 3\zet,n} y^{m+H} z_2^{n+L} \\
\frac{\Gamma(1+2(m+H))}{\Gamma(1+\frac13(m+H))^2
\Gamma(1+m+H) \Gamma(1-\frac13(m+H)+n+L)^2\Gamma(1+m+H-2(n+L))}
\end{multline}
and the solution of \eqref{goodpower} given above can be understood from the representation
\begin{equation}
\calw (y,z_2) = 3\bigl[\frac 12\del_H^2 \tilde\varpi(y,z_2;0,0) +  \tau(y,z_2)\bigr]
\end{equation}
where
\begin{equation}
\tau(y,z_2) = \frac{4\pi^2}{27} \sum_{m\notin 3\zet,n}
\frac{\Gamma(1+2m)}{\Gamma(1+\frac m3)^2\Gamma(1+m)
\Gamma(1-\frac m3+n)^2\Gamma(1+m-2n)} y^m z_2^n
\end{equation}
Indeed, by construction,
\begin{multline}
9\call_2\tilde\varpi(y,z_2;H,L) = \sum_{m\in 3\zet}  y^{m+H} z_2^L \\ 
 \frac{(H+m-3L)^2\Gamma(1+2(m+H))}{\Gamma(1+\frac 13(m+H))^2
\Gamma(1+m+H)\Gamma(1-\frac 13(m+H)+L)^2\Gamma(1+m+H-2L)} 
\end{multline}
When acting with $\del_H^2$, and restricting to $H=L=0$, the terms at $m\neq 0$ would 
vanish because of the appearance of $\Gamma(1-\frac m3)^2$ in the denominator, {\it unless} 
both derivatives act on that factor, to yield
\begin{equation}
\sum_{\topa{m\in 3\zet}{m>0}} \frac{m^2 \Gamma(1+2m) \Gamma(\frac m3)^2 \cos(\pi \frac m3)^2}
{9\Gamma(1+\frac m3)^2 \Gamma(1+m)^2} y^m
=
\sum_{\topa{m\in 3\zet}{m>0}}\frac{\Gamma(1+2m)}{\Gamma(1+m)^2} y^m
\end{equation}
On the other hand, the term at $m=0$ gives a non-zero contribution only when both derivatives
act on $(H-3L)^2$. Also,
\begin{equation}
\begin{split}
9\call_2\tau(y,z_2) &= \frac{4\pi^2}{27}
\sum_{m\notin 3\zet} \frac{m^2 \Gamma(1+2m)}{\Gamma(1+\frac m3)^2 \Gamma(1+m)
\Gamma(1-\frac m3)^2\Gamma(1+m)} y^m \\
&=\sum_{m\notin 3\zet} \frac{\Gamma(1+2m)}{\Gamma(1+m)^2} y^m
\end{split}
\end{equation}
In combination, we obtain indeed the inhomogeneity in the form 
\eqref{expanding}. A similar computation verifies the equation for $\call_1$. 

Note that we may reinstate the discrete labels on the superpotential via
\begin{equation}
\calw_{(\ph,\pm)}(y,z_2) = \pm \calw_{(1,+)}(\ph y,z_2)
\end{equation}

\subsection{Mirror map and instanton sum}

Most of the rest of the paper is devoted to verifying that the solution of the
inhomogeneous Picard-Fuchs equation has a consistent interpretation as a global
holomorphic object over the entire moduli space. As an important first check, we will 
here show that the expansion around $y=z_2=0$ satisfies Ooguri-Vafa integrality, \ie,
has a consistent interpretation as counting domainwall degeneracies.

The main step is to understand the mirror map. For this, note that the generating function 
\eqref{withwhich} can of course also be used to express the solutions of the homogeneous 
equation. In particular, the regular and simple logarithmic solutions are
\begin{equation}
\begin{split}
\varpi_0 &= \tilde\varpi(y,z_2;0,0) \\
\varpi_h &= \del_H\tilde\varpi(y,z_2;0,0)= \frac 13 \bigl(\varpi_{h_1} - \varpi_{h_2}\bigr) \\ 
\varpi_l &= \del_L\tilde\varpi(y,z_2;0,0)= \varpi_{h_2} \,.
\end{split}
\end{equation}
where the relations to \eqref{soltwo} are dictated by \eqref{curverels}. By inspecting 
\eqref{someof}, we may anticipate from our discussion in section \ref{phasetransition} that 
the tension of a supersymmetric domainwall between vacua labelled $\ph$ and
$\ee^{2\pi\ii/3}\ph$ behaves to leading order as $\calt=\calw_{(\ee^{2\pi\ii/3}\ph,\pm)}
-\calw_{(\ph,\pm)} \sim \varpi_h + \cdots$. 
Translated into the A-model, this means that the large volume geometry must
admit a domainwall with classical tension
\begin{equation}
\calt_{\rm class} = \frac 13 (t_1-t_2) \equiv s
\end{equation}
Following \cite{oova}, one may pose the problem to count the degeneracy of such
domainwalls, and this information should be contained in the B-model superpotential 
$\calw_{(\ph,\pm)}$. As in the previous compact examples, and in agreement with the 
structure found in non-compact examples \cite{agva,akv,lmw}, the prescription is 
to {\it expand the superpotential with open string instanton corrections in terms 
of the classical domainwall tension corrected only by closed string instantons}. 
In the problem at hand, we introduce
\begin{equation}
\begin{split}
o & = \exp\Bigl(\frac{\varpi_h}{\varpi_0}\Bigr) = \ee^{2\pi\ii s} \\
q &= \exp\Bigl(\frac{\varpi_l}{\varpi_0}\Bigr) = \ee^{2\pi\ii t} = q_2 
\end{split}
\end{equation}
\begin{table}[t]
\begin{center}
\begin{tabular}{|l|lllllll|}
\hline
$m\setminus n$ & 0 & 1 & 2 & 3 & 4 & 5 & 6 \\\hline
0 & 0 & 1 & 0 & 0 & 0 & 0 & 0 \\
1 & 6 & 6 & 0 & 0 & 0 & 0 & 0 \\
2 & 3 & 90 & 3 & 0 & 0 & 0 & 0 \\
3 & 6 & -236 & 1012 & 6 & 0 & 0 & 0 \\
4 & 12 & -258 & 2934 & -258 & 12 & 0 & 0 \\
5 & 30 & -540 & 11016 & 11016 & -540 & 30 & 0 \\
6 & 75 & -1388 & -44274 & 348 774 & 179478 & -1388 & 75\\\hline
\end{tabular}
\end{center}
\caption{Open BPS invariant of two-parameter model.}
\label{mainresult}
\end{table}
The A-model expansion now takes the following form (we have omitted the
constant term \eqref{rationalize})
\begin{equation}
\eqlabel{omitted}
\hat\calw_{(\ph,\pm)} = \frac{\calw_{(\ph,\pm)}}{\varpi_0} =
\pm\Bigl[\frac 32 (\log \ph o)^2 + \sum_{\topa{(m,n)\neq (0,0)}{k>0}} \frac{N_{m,n}}{k^2} 
(\ph o)^{k m} q^{k n}
\Bigr]
\end{equation}
The $N_{m,n}$ are indeed integer. We display the first few in table \ref{mainresult}. 
Note that the invariants are symmetric under $n\to m-n$, except when $m$ is a multiple 
of $3$. We may improve on this after recognizing that the asymmetry is in fact a remnant 
of closed string instantons. More precisely, we have for $m\in 3\zet$ (and $m\ge n$)
\begin{equation}
N_{m,n}-N_{m,m-n} = \frac{(2n-m)}2 G_{\frac m3,n-\frac m3}
\end{equation}
where $G_{i,j}$ are the standard closed string BPS invariants computed in 
\cite{cdfkm}. (The $G_{i,j}$ are themselves symmetric under $j\to i-j$ and vanish outside
of $0\le j\le i$.)
Expressed in terms of the standard variables \eqref{standardmir}, the 
combination $(2n-m) G_{i,j}$ enters into the closed string instanton
corrections of a particular 4-cycle tension
\begin{equation}
\eqlabel{4cyc}
2\hat\varpi_{H_2} - \hat\varpi_{H_1} = 
-2t_1t_2 + \sum_{i,j,k} \frac{(2i-j)G_{i,j}}{k^2} q_1^{ki} q_2^{k j}
\end{equation}
So after transforming to the new variables, the ``balanced'' superpotential
\begin{equation}
\eqlabel{impsupo}
\tilde \calw = \calw-\frac 14(2\varpi_L-\varpi_H)=
\calw + \frac 32\del_H\del_L\tilde \varpi+ \frac 12\del_L^2 \tilde\varpi
\end{equation}
has an A-model expansion
\begin{equation}
\frac 32 s^2 + \frac 32 st + \frac 12 t^2 + \sum_{m,n,k} 
\frac{\tilde N_{m,n}}{k^2} o^{km} q^{k n}
\end{equation}
with invariants $\tilde N_{m,n}$ that are now symmetric under $n\to m-n$,
see table \ref{mainresult2}. One might try to corroborate this symmetry 
and the modification \eqref{impsupo} in terms of monodromy calculations.
\begin{table}[t]
\begin{center}
\begin{tabular}{|l|lllllll|}
\hline
$m\setminus n$ & 0 & 1 & 2 & 3 & 4 & 5 & 6 \\\hline
0 & 0 & 0 & 0 & 0 & 0 & 0 & 0 \\
1 & 6 & 6 & 0 & 0 & 0 & 0 & 0 \\
2 & 3 & 90 & 3 & 0 & 0 & 0 & 0 \\
3 & 6 & 388 & 388 & 6 & 0 & 0 & 0 \\
4 & 12 & -258 & 2934 & -258 & 12 & 0 & 0 \\
5 & 30 & -540 & 11016 & 11016 & -540 & 30 & 0 \\
6 & 75 & -1388 & 67602 & 348 774 & 67602 & -1388 & 75\\\hline
\end{tabular}
\end{center}
\caption{``Balanced'' invariants with manifest symmetry under $m\to n-m$.}
\label{mainresult2}
\end{table}

\def\smallfrac#1#2{{\textstyle\frac{#1}{#2}}}

\section{Open-Closed Moduli Space of Two-parameter Model}
\label{modulispace}

For the subsequent computations, it is useful to have a good global picture of the 
combined open-closed moduli space. 

Following \cite{cdfkm}, one begins by noting that the parameter space of \eqref{poly2par}, 
spanned by $(\psi,\phi)$, is subject to a $\zet_{12}$ quotient generated by 
$(\psi,\phi)\to (\ee^{2\pi\ii/12}\psi,-\phi)$, since this action can be undone
by the change of coordinate $x_1\to \ee^{-2\pi\ii/12} x_1$. The invariant combinations
are
\begin{equation}
\xi = \psi^{12}\,,\quad \upsilon=\psi^6\phi \,,\quad \zeta=\phi^2 \,,
\end{equation}
now subject to the relation
\begin{equation}
\eqlabel{subject}
\xi\zeta=\upsilon^2 \,.
\end{equation}
A first model of the compactified moduli space is obtained by viewing $(\xi,\upsilon,\zeta)$
as inhomogeneous coordinates in the patch $\tau=1$ of a copy of $\projective^3$. 
In this compactification, there are four special loci along which the family acquires various 
singularities. These are
\\
\noindent
$\bullet$ the conifold locus $C_{\rm con}= \{\xi+ 2\upsilon+\zeta=\tau\}$,
(In $\psi,\phi$ space, it is the locus $(\phi+\psi^6)^2=1$.)
\\
\noindent
$\bullet$ the locus $C_1=\{\zeta=\tau\}$ (compactification of $\phi^2=1$),
\\
\noindent
$\bullet$ the limit of large $\psi,\phi$, $C_{\infty}=\{\tau=0\}$,
\\
\noindent
$\bullet$ the locus of enhanced symmetry $C_0=\{\xi=0,\upsilon=0\}$ ($\psi=0$).

\begin{figure}[t]
\begin{center}
\psfrag{C1}{$C_1$}
\psfrag{C0}{$C_0$}
\psfrag{Cd}{$C_d$}
\psfrag{Ci}{$C_\infty$}
\psfrag{Cc}{$C_{\rm con}$}
\epsfig{file=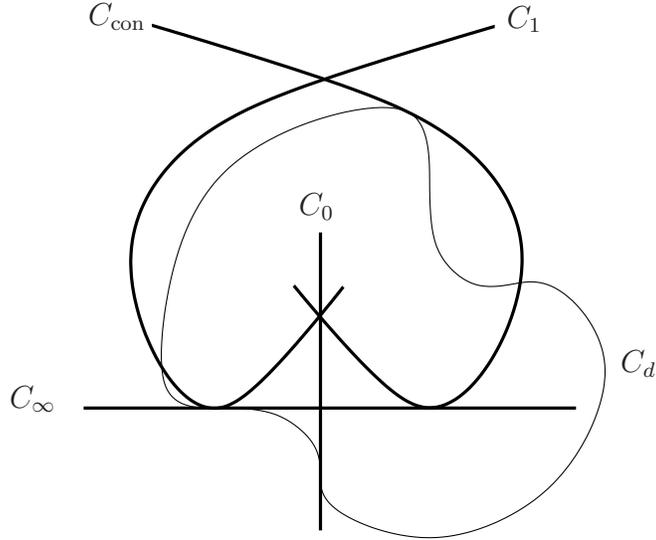,height=7cm}
\end{center}
\caption{The compactified moduli space before blowups. The four divisors of closed string
origin, depicted with a thicker line, include the large volume divisor $C_\infty$, the
conifold locus $C_{\rm con}$, the enhanced symmetry locus $C_0=\{\psi=0\}$ and the locus 
$C_1=\{\phi^2=1\}$. The thinner line represents the open string discriminant $C_d$. The 
orbifold point lies at the intersection of $C_0$ and $C_d$.}
\label{beforeblowup}
\end{figure}

These divisors meet at various points. Not all of these intersections are
transverse, so one needs to blow up to obtain a good compactification. In
practice, this is accomplished by working in the appropriate local 
coordinates. For example, the large complex structure point, which is (here 
uniquely) characterized by maximal unipotent monodromy, is hidden at the double 
intersection of $C_1$ and $C_\infty$. In the patch $\xi=1$, where we can
eliminate $\zeta$ using \eqref{subject}, this is the point $\tau=1/\psi^{12}=0$, 
$\upsilon=\phi/\psi^6=0$. Blowing up once introduces the coordinate 
$\alpha=\tau/\upsilon=1/(\phi\psi^6)$ on the divisor called $D_{(-1,-1)}$ in
\cite{cdfkm}. $D_{(-1,-1)}$, $C_1$ and $C_\infty$ now all meet at
$\alpha=\upsilon=0$. A second blowup of this point inserts the divisor
$D_{(-1,0)}$, with coordinate $\beta=\alpha/\upsilon=1/\phi^2$ that
now intersects $C_\infty$ transversely. The coordinates $\upsilon\sim z_1$,
$\beta\sim z_2$ are precisely those appropriate for the closed string
mirror map in the previous section.

Now let us add the D-brane. At a generic point of the moduli space, we are considering 
$6$ disjoint curves $C_{(\ph,\pm)}$ which we think of physically as representing
6 different vacua of some $\caln=1$ theory in 4 dimensions. The combined open-closed
moduli space is thus a six-fold cover of the closed moduli. However, there
are various places where some of these vacua come together, and/or are permuted
in various ways under monodromy. The merging of vacua is accompanied by new
light physical degrees of freedom, but is not necessary for the occurrence of 
monodromy. Since the monodromies should be consistent with the symmetry of the 
problem, there are only 3 non-trivial possibilities: A $\zet_2$, a $\zet_3$, or a 
$\zet_6\cong\zet_2\times\zet_3$ rotation of the vacua. 

\begin{figure}[t]
\begin{center}
\psfrag{C1}{$C_1$}
\psfrag{C0}{$C_0$}
\psfrag{Cd}{$C_d$}
\psfrag{Ci}{$C_\infty$}
\psfrag{Cc}{$C_{\rm con}$}
\psfrag{E1}{$E_1$}
\psfrag{E0}{$E_0$}
\psfrag{E2}{$E_2$}
\psfrag{Dm1}{$D_{(-1,0)}$}
\psfrag{Dm11}{$D_{(-1,-1)}$}
\psfrag{D01}{$D_{(0,-1)}$}
\epsfig{file=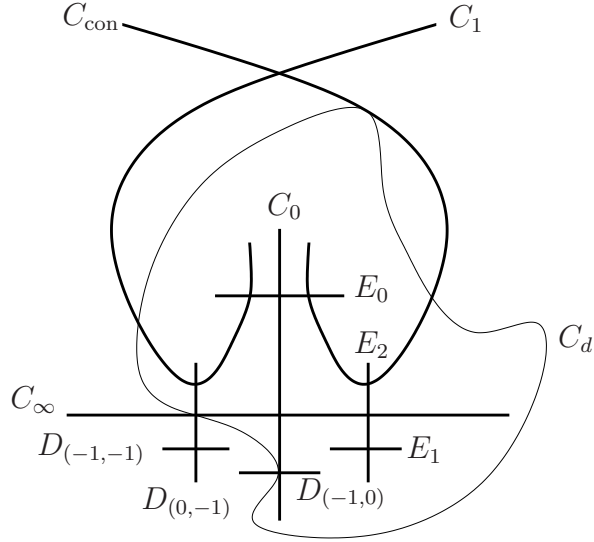,height=7cm}
\end{center}
\caption{The moduli space blown up according to \cite{cdfkm} such that the closed string 
boundary divisor has normal crossings. The large volume point lies at the intersection 
of $C_\infty=\{z_1=0\}$ with $D_{(0,-1)}=\{z_2=0\}$, and is also met by $C_d$.}
\label{closedblowups}
\end{figure}

It is rather straightforward to identify some of these loci from the right side of 
\eqref{fornow}. For this, we note that $y^3=z_1/z_2\sim \phi^3/\psi^6$ is the coordinate 
on an additional blowup of the moduli space at the large complex structure point. We call 
this exceptional divisor $D_d$. Taking a third root produces a threefold cover branched 
at $y=0$. We can then see
\nxt a $\zet_3$ monodromy, acting on $(\ph,\pm)\to(\ee^{2\pi\ii/3}\ph,\pm)$, 
at $y^3=0$,
\nxt a $\zet_2$ monodromy, acting as $(\ph,\pm)\to (\ph,\mp)$, at the locus
$(4y)^3=1$. We will denote this divisor by $C_d$. In the homogeneous
coordinates, it is the locus
\begin{equation}
C_d=\{\xi\zeta-\upsilon^2=0,\; 27\upsilon\tau+4\zeta^2=0,\; 729\xi\tau^2-16\zeta^3=0\}
\end{equation}
\fnxt a $\zet_6$ monodromy, acting as $(\ph,\pm)\to (\ee^{2\pi\ii/3}\ph,\mp)$
at $x\equiv y^{-1}=0$.  

\begin{figure}
\begin{center}
\psfrag{Cd}{$C_d$}
\psfrag{Ci}{$C_\infty$}
\psfrag{D01}{$D_{(0,-1)}$}
\psfrag{Dd}{$D_d$}
\epsfig{file=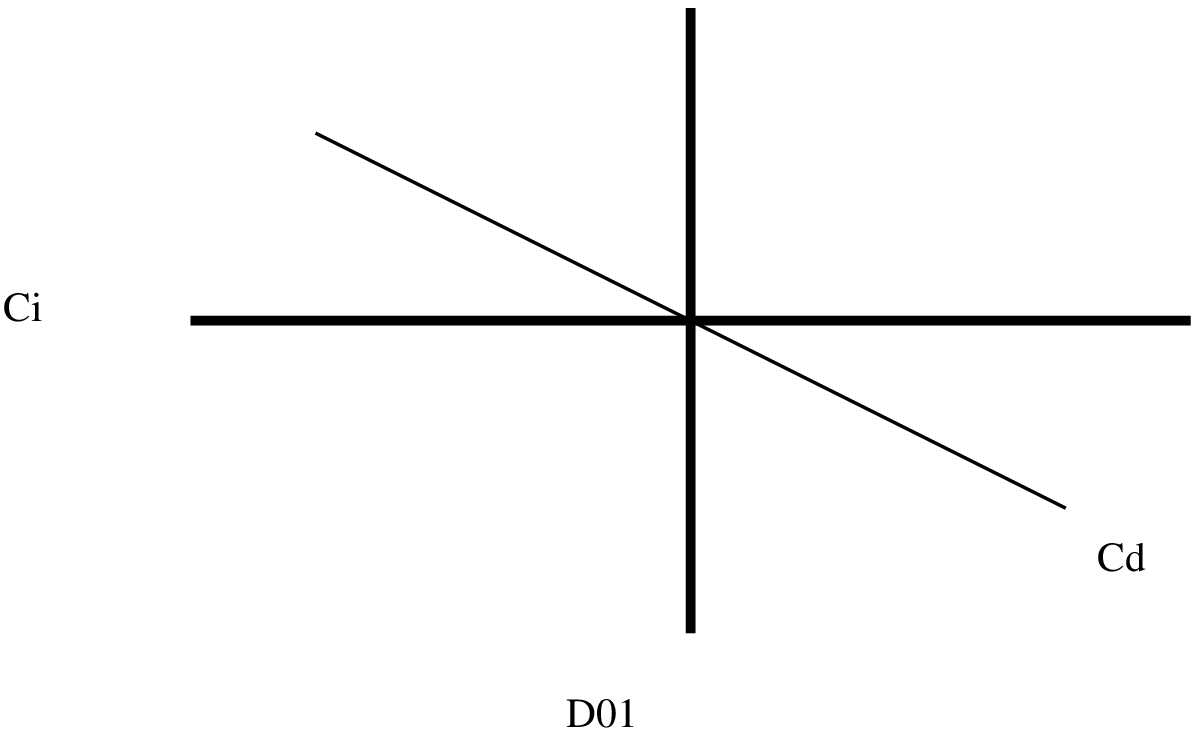,height=3cm}
\qquad\qquad
\epsfig{file=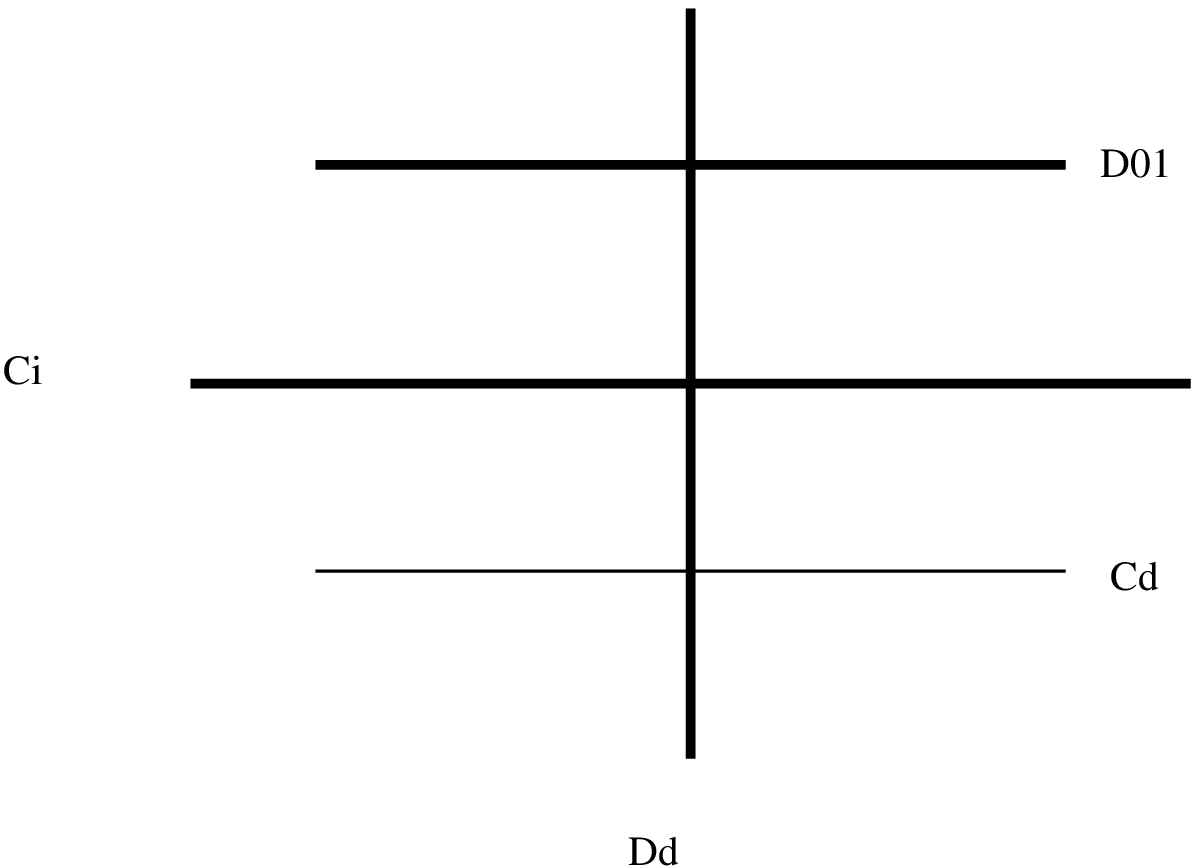,height=3cm}
\end{center}
\caption{Close up of large volume point, with additional blowup depicted on the right.
The coordinate on the exceptional divisor $D_d$ is denoted $y^3=1/x^3\sim\frac{z_1}{z_2}$
in the text.}
\label{closeup}
\end{figure}

It is straightforward to analyze the intersections of $C_d$ with the loci $C_0$, 
$C_1$, $C_{\rm con}$, and $C_\infty$ listed above. We have depicted the result 
in Fig.\ \ref{beforeblowup}, and the state of affairs after the blowups of
\cite{cdfkm} in Fig.\ \ref{closedblowups}. We will investigate in more detail
the blowup of the large complex structure point, sketched in Fig.\ \ref{closeup}, in 
section \ref{phasetransition}. It is likely that the other special points on $C_d$
might also harbor interesting physical effects.

\section{Restriction to One-parameter Model}
\label{restrict}

It is of interest to pay some more attention to the locus $\phi=\pm 1$ ($z_2=1/4$) 
of the closed string moduli space. For these values of the parameters \cite{cdfkm}, 
the two-parameter model $\projective_{11226}[12]$ becomes birationally equivalent 
to the one-parameter model $\projective_{111113}[2,6]$. This entails a certain
relationship between the quantum geometries of the two models that provides
a useful check on the calculations.

In our context, we have another reason to look at these relations. As noted
in the introduction, $\projective_{111113}[2,6]$, although its periods are
hypergeometric, does not seem to possess a brane whose superpotential
can be obtained by specializing $\varpi(z;H)$ to particular simple values of $H$.
(Most noticeably, $\Xi(1/2)= 16/\pi^3$.) By restricting the superpotential of
the two-parameter model, we are able to produce a sensible extension of the
Picard-Fuchs equation also in that case.

The basic relation between the two models is the identity of
fundamental periods
\begin{equation}
\varpi^{(1)}_0(z) =\varpi^{(1)}(z;0) = \varpi_0(2z^{1/3},1/4)
\end{equation}
where $\varpi^{(1)}(z;H)$ is the appropriate hypergeometric generating
function. The relation for the other periods is less direct, although we 
still have
\begin{equation}
\varpi_{1}^{(1)}(z) = 3  \varpi_h(2 z^{1/3},1/4) + \frac 32 
\varpi_l(2 z^{1/3},1/4)
\end{equation}
It is then natural to study the restriction of the superpotential
$\calw_{(\ph,\pm)}$. Since the twice logarithmic periods of the two-parameter
model do not restrict to a period of the one-parameter model, we should
allow for a modification of $\calw$ by a rational period. The best result is 
obtained for the combination
\begin{equation}
\calw^{(1)}(z) = \bigr(\calw + \frac 13\varpi_H -\frac12{\varpi_L} \bigr)(2z^{1/3},1/4)
= \bigr(\tilde\calw + \frac{1}{12}\varpi_H\bigr)(2 z^{1/3},1/4)
\end{equation}
Applying the Picard-Fuchs operator of $\projective_{111113}[2,6]$,
\begin{equation}
\call^{(1)}= \theta^4 - \frac{16}3 z (6\theta+1)(6\theta+3)^2(6\theta+5)
\end{equation}
gives the remarkable inhomogeneity
\begin{equation}
\eqlabel{remarkable}
\begin{split}
\call^{(1)} \calw^{(1)}(z) &= \sum_{n=1}^\infty \frac{2^{n+1}\Gamma(2n)}{27\Gamma(n)^2}
(10n-9)z^n \\
&= \frac{4}{27} \frac{z^{1/3}+112 z^{2/3}}{(1-8z^{1/3})^{5/2}} 
\end{split}
\end{equation}
One should now expect that after inserting the mirror map of the one-parameter model, 
$\calw^{(1)}$ has an integer BPS expansion that moreover is related to the expansion
of the two-parameter model in a simple way. Recall that in the closed string sector,
one has the sum rule \cite{cdfkm}
\begin{equation}
\eqlabel{srule}
\sum_{j=0}^i G_{i,j} = G^{(1)}_i
\end{equation}
where $G^{(1)}_i$ are the invariants appearing in the expansion
($p=\exp(\varpi^{(1)}_1/\varpi^{(1)}_0)$)
\begin{equation}
\hat\varpi^{(1)}_2 = 2(\log p)^2 + \sum_{i,k} \frac{iG^{(1)}_i}{k^2} p^{ki}
\end{equation}
For $\calw^{(1)}$, such a relation only emerges after a modification 
analogous to \eqref{impsupo}.
\begin{equation}
\eqlabel{analogous}
\tilde\calw^{(1)} = \calw^{(1)} - \frac 14 \varpi^{(1)}_2
\end{equation}
has the A-model expansion
\begin{equation}
\frac 16(\log p)^2 + \sum_{m} \frac{N^{(1)}_m}{k^2} p^{k m/3} 
\end{equation}
These invariants now satisfy
\begin{equation}
\eqlabel{sumrule}
N^{(1)}_m = \sum_{n=0}^m N_{m,n}
\end{equation}
while the first few of them are
\begin{equation}
N_1=12\,,\; N_2= 96\,,\; N_3=788\,,\; N_4=2442\,,\; N_5=21012\,,\;
N_6=481352\,,\;\ldots
\end{equation}
It will be interesting to verify the relation \eqref{sumrule} by an analysis of the 
A-model geometry. It is also worthwhile to use the rational mapping \footnote{I am 
grateful to Dave Morrison and Sheldon Katz for recovering this birational 
equivalence from \cite{cdfkm}.} between the B-model geometries to obtain curves 
representative of the superpotential in the one-parameter model, and to then derive 
the inhomogeneity \eqref{remarkable} by the Griffiths-Dwork method. This computation 
is, however, somewhat uncertain because the B-model geometry of the one-parameter model is 
singular (for generic values of the parameter) precisely at the points where the 
residues should be localized. If the modifications \eqref{impsupo}, \eqref{analogous}
seem a bit ad hoc, we point out that the actual comparison of closed string invariants
\cite{cdfkm} takes place at the level of the Yukawa couplings. The open string analogue
would be a comparison of Griffiths infinitesimal invariants \cite{openbcov}.

\section{D-brane Phase Transitions at Large Volume}
\label{phasetransition}

In this section, we study and interpret the analytic properties of the solutions of the
differential equation
\begin{equation}
\eqlabel{diffeq}
\theta_y^2 V(y) = \frac{1}{(1-4y)^{1/2}}
\end{equation}
where $\theta_y = y\frac{d}{dy}$. This equation is the restriction of the
inhomogeneous Picard-Fuchs equation \eqref{goodpower} to the locus $z_2=0$. The
complex $y$-plane is a three-fold cover of the divisor $D_d$ that we introduced
in section \ref{modulispace}. There are three special points. The intersection
with $C_\infty$ corresponds to $y=0$, the intersection with $D_{(0,-1)}$ is
the point $y=\infty$, and the intersection with $C_{d}$ occurs at $y=\frac 14$.

\subsection{Analytic continuation via dilogarithm}

It is natural to begin and fix boundary condition at $4y=1$. In terms of the
variable
\begin{equation}
w= 1-4y
\end{equation}
the differential equation \eqref{diffeq} is
\begin{equation}
(1-w)\del_w (1-w)\del_w V = w^{-1/2}
\end{equation}
and we pick the solution of the inhomogeneous equation that vanishes at $w=0$.
This solution can be written as
\begin{equation}
\eqlabel{T0}
V(w) = 
\frac 12 \Bigl(\log\frac{1-\sqrt{w}}{2}\Bigr)^2
-\frac 12 \Bigl(\log\frac{1+\sqrt{w}}{2}\Bigr)^2
+\Li_2 \frac {1-\sqrt{w}}{2} - \Li_2\frac{1+\sqrt{w}}{2}
\end{equation}
Here $\Li_2(z)$ is the dilogarithm. It can be defined by its power-series
expansion around $z=0$
\begin{equation}
\Li_2(z) =\sum_{n=1}^\infty\frac {z^n}{n^2} 
\end{equation}
and analytically continued through the complex plane via the integral
\begin{equation}
\eqlabel{view}
\Li_2 (z) = -\int \frac{\log(1-z)}z \; dz
\end{equation}
with a branch cut from $z=1$ to $z=\infty$.

There is now little work for us to do if we are willing to refer to some standard 
properties of the dilogarithm. Using the transformation
\begin{equation}
\Li_2(z) = -\Li_2(1-z) + \frac{\pi^2}6 - \log z\log (1-z)
\end{equation}
we obtain the representation ($y=(1-w)/4$),
\begin{equation}
V =  \frac 12 (\log y)^2 - \frac{\pi^2}6 -\Bigl(\log\frac{1+\sqrt{1-4y}}2\Bigr)
+  2\Li_2 \frac{1-\sqrt{1-4y}}2 
\end{equation}
which gives the expansion around $y=0$,
\begin{equation}
\eqlabel{exp1}
V =\frac{(\log y)^2}{2} - \frac{\pi^2}{6} + \sum_{m=1}^\infty \frac{\Gamma(1+2m)}
{\Gamma(1+m)^2 m^2} y^m
\end{equation}
Note that this agrees with \eqref{z20sol} up to a constant, which will play an
important role below. On the other hand, applying the identity (valid under continuation 
through the upper half plane)
\begin{equation}
\Li_2(1/z) = - \Li_2(z) - \frac{\pi^2}{6} - \frac 12\bigl(\log(-z)\bigr)^2
\end{equation}
we obtain (with $x=1/y$),
\begin{equation}
V= \ii\pi\ln x- \frac 12 \Bigl(\log\frac{\sqrt{x^2-4x}+x}{\sqrt{x^2-4x}-x}\Bigr)^2
-2 \Li_2\bigl(\frac{\sqrt{x^2-4x}+x}2\bigr)
\end{equation}
which gives the expansion around $x=0$
\begin{equation}
\eqlabel{exp2}
V = \ii\pi\log x - \ii \sum_{m=0}^\infty \frac{\Gamma(1+2m)}
{\Gamma(1+m)^2 (m+\frac12)^2 2^{4m+1}} x^{m+1/2}
\end{equation}
As a consistency check, we note the transformations of $V$ under the monodromies
around the three points
\begin{equation}
\begin{split}
y\to \ee^{2\pi\ii} y:& \; V\to V+2\pi\ii\log y + \frac{(2\pi\ii)^2}2 \\
z \to \ee^{2\pi\ii} z:&\; V\to -V \\
x\to \ee^{2\pi\ii} x: &\; V\to -V +2\pi\ii\log x + \frac{(2\pi\ii)^2}2
\end{split}
\end{equation}
which indeed compose correctly.

\subsection{Interpretation}

The information gathered so far is sufficient to predict some non-trivial properties of the 
A-model geometry that is mirror to the configuration of holomorphic curves introduced in 
section \ref{twopar}. Classically, vacuum configurations of A-branes are described by Lagrangian 
submanifolds $L\subset X$ equipped with flat bundles. There are however, quantum corrections
from worldsheet instantons wrapping holomorphic disks \cite{kklm} that affect the vacuum 
structure in a qualitative way. Since instanton corrections die out when the K\"ahler
moduli are taken to infinity, but are present for any finite value, the vacuum structure of
A-branes can change in a discontinuous way in the large volume limit. This complication is at 
the heart of understanding mirror symmetry for open strings in an invariant way.

In our problem, a minimum requirement is that the A-brane should have 6 supersymmetric vacua 
for finite values of the K\"ahler moduli. We can use the structure of the superpotential, or
more precisely the spectrum of domainwall tensions, to deduce what the corresponding
classical configurations should look like.

Recall that in section \ref{twopar} we introduced the discrete labels $(\ph,\pm)$, where $\ph$
is a third root of unity. We will presently see that because of the mixing between open 
strings and Ramond-Ramond flux degrees of freedom, it is actually more efficient to resort
to $\zet$-valued labels. Thus we write 
\begin{equation}
\ph = \ee^{2\pi\ii a/3} \,,\qquad
\pm = \ee^{2\pi\ii b/2}
\end{equation}
and take $a,b\in\zet$. We denote the tension of supersymmetric domainwalls between adjacent 
vacua as
\begin{equation}
\begin{split}
\calt_{2,(a,b)} &= \calw_{(a+1,b)} - \calw_{(a,b)} \\
\calt_{3,(a,b)} &= \calw_{(a,b+1)} - \calw_{(a,b)} 
\end{split}
\end{equation}
Up to closed string periods, we can deduce the asymptotic form of $\calt_2$, $\calt_3$ from 
our computations in section \ref{twopar}. In the large volume expansion in the A-model, 
for $t_1\gg t_2$ we find from \eqref{omitted} (we use $s=(t_1-t_2)/3$, and absorb all 
factors of $(2\pi\ii)$)
\begin{equation}
\eqlabel{comparing}
\begin{split}
\calt_{2,(a,b)} &= (-1)^b \frac 13 (t_1-t_2+a)^2 + \cdots \\
\calt_{3,(a,b)} &= (-1)^b \frac 13 (t_1-t_2+a) + (-1)^b\frac 16+ \cdots
\end{split}
\end{equation}
We now give an interpretation of this structure from the A-model point of view, after
which we will see that there are other subleading corrections to \eqref{comparing}. To fix 
ideas somewhat more generally, we consider, in type IIA string theory compactified 
on a Calabi-Yau $X$, a D6-brane wrapped on a Lagrangian submanifold $L\subset X$. 
If we assume that $L$ is classically rigid, \ie, $b_1(L)=0$, possible choices of 
the gauge field are distinguished by the value of a discrete Wilson line $w\in 
H_1(L,\zet)\cong H^2(L,\zet)$, or equivalently a first Chern class. 

There can now be two types of domainwalls. We can change the Lagrangian submanifold
to a homologous Lagrangian $L'$, or we can change the value of the gauge field to $w'$. 
The first type of transition (say around $0=x_3\in\reals\subset\reals^{3,1}$) is represented in 
space-time by a supersymmetric 4-cycle in $X\times\reals$ that asymptotes to $L$ or
$L'$ for $x_3\to \pm \infty$, respectively. In the Calabi-Yau, the three-cycle sweeps
out a four chain $\Gamma_4$, with $\del\Gamma_4= L'-L$. The classical BPS tension of 
such a domainwall is in the large volume limit given by
\begin{equation}
\int_{\Gamma_4} J\wedge J
\end{equation}
where $J$ is the complexified K\"ahler form. To change the value of the magnetic
flux $w$, we pick a relative 2-cycle $\Gamma_2\in H_2(X,L)$ that ends on $L$ in a
one-cycle equivalent to $w'-w$. We then wrap a D4-brane on $\Gamma_2 \times \{x_3=0\}$. 
Its tension is classically given by
\begin{equation}
\int_{\Gamma_2} J
\end{equation}
In our case, it is easy to distinguish the two types from \eqref{comparing}. Since 
$J\sim t_i$, we see from the scaling behavior that $\calt_2$ must correspond to a 
domainwall changing the Lagrangian submanifold, whereas $\calt_3$ corresponds to a change of 
magnetic flux. This is consistent with having two rigid Lagrangians each with 
fundamental group $\zet_3$.

But there are additional constraints that we have to take into account. From the
A-model perspective, large volume monodromies, \ie, changes of $t_i$ by integers, must 
have a consistent interpretation as a symmetry of brane vacua and domainwall spectrum.
As pointed out in \cite{open}, the apparent lack of periodicity of \eqref{comparing}
in $a$ can be compensated by a non-trivial action on the Ramond-Ramond fluxes.
Specifically, under $a\to a+3$, $b\to b+2$, the domainwalls must return
to themselves up to an integral closed string period. The possibilities here include
$t_1$, $t_2$, and $1$, interpreted as changing the RR 4 and 6-form flux respectively.
Similar reasoning implies that $\calt_{3,(a,b)}+ \calt_{3,(a+1,b)} + \calt_{3,(a+2,b)}$
and $\calt_{2,(a,b)}+\calt_{2,(a,b+1)}$ should also be integral closed string
periods. (These requirements are equivalent to the superpotential having integral
monodromy.)

Another constraint comes from the B-model, and the fact that the curves $C_{(\ph,+)}$
and $C_{(\ph,-)}$ merge together when $4\ph y=1$. This means that $\calt_{2}$ is a 
tensionless domainwall at the open string discriminant $C_d$. The calculations in the 
previous subsection then imply that on $D_d$, for $t_1\gg t_2$, $\calt_2$ should 
asymptote to 
\begin{equation}
6V\sim \frac 13(t_1-t_2)^2 + \frac 14
\end{equation}
Implementing all these constraints, we find that the correct asymptotic behavior
of the domainwalls is given by
\begin{equation}
\eqlabel{final}
\begin{split}
\calt_{2,(a,b)} &= (-1)^b \frac 13 (t_1-t_2+a)^2 +(-1)^b \frac 14 + \cdots \\
\calt_{3,(a,b)} &= (-1)^b \frac 13 (t_1-t_2+a) +\frac{(-1)^b+1}6 +  \cdots
\end{split}
\end{equation}
where now the dots only contain worldsheet instanton corrections that are determined
from the solution of the differential equation.  For completeness, we note that the
structure \eqref{final} can be derived from the modified superpotential
\begin{equation}
\eqlabel{finalW}
\calw_{(a,b)} = (-1)^b\frac 16(t_1-t_2+a)^2 + \frac 16 (t_1-t_2+a) + (-1)^b \frac 18 
+ \cdots
\end{equation}
Since this differs from \eqref{omitted} only by integral periods, the modification does 
not interfere with the inhomogeneous Picard-Fuchs equation.

Using mirror symmetry, and the differential equations, we can analytically continue these
expressions through the entire moduli space. This could be used to check integrality of
the monodromies around the small volume phases of the closed string geometry.

For the moment, we will however restrict attention to the behavior under analytic
continuation on the exceptional large volume divisor $D_d$. Notice that as we move
towards $C_d$, $\calt_2$ decreases. This means that under this deformation of 
symplectic structure of $X$, the two Lagrangians classically approach each other
(this is another indication that the Lagrangians should not be real slices). 
Simultaneously however, worldsheet instanton corrections become strong, spoiling
the geometric picture. Another semi-classical regime emerges for $t_2\gg t_1$,
\ie, close to $D_{(0,-1)}$. What can we say about the Lagrangian geometry in 
this phase? From \eqref{exp2}, we find the asymptotic behavior in this regime
\begin{equation}
\eqlabel{remarkab}
\begin{split}
\tilde \calt_{2,(a,b)} &= (-1)^b (t_2-t_1) +\cdots  \\
\tilde \calt_{3,(a,b)} &= (-1)^b \frac 12 + \frac 16 + \cdots
\end{split}
\end{equation}
where the dots again only contain non-perturbative corrections. We see from these
expressions that there should be only a single Lagrangian submanifold $\tilde L$
relevant in this phase. The domainwall $\calt_2$ must correspond to a holomorphic 
disk ending on $\tilde L$. Moreover, the domainwall $\calt_3$ classically has 
vanishing tension (in general, the constant non-integral terms in the above 
superpotentials are expected to arise from {\it perturbative} worldsheet 
corrections). Thus it appears that in this regime, although all closed string 
worldsheet instanton corrections are small, the vacuum structure on $\tilde L$ 
cannot be understood classically. 

There is a hopefully more invariant way to characterize the occurrence of this
phase transition. The asymptotic form of $\calt_3$ for $t_1\gg t_2$ implies 
that the Lagrangians $L$, $L'$ admit the boundary of holomorphic disks whose 
symplectic area scales as $(t_1-t_2)/3$. (This is precisely the variable in which 
we have expanded the superpotential in section \ref{twopar}. Note that the notion 
is well-defined since when $H_1(L)$ is torsion we can invert the map $H_2(X)\to 
H_2(X,L)$ over the rationals.) Such holomorphic disks then introduce additional 
walls in the K\"ahler cone, across which the disks undergo a transition reminiscent 
of a ``flop''. (The Lagrangian in the other phase does not seem to have 
domainwalls scaling as $(t_2-t_1)/3$. The instanton corrections suggest
that the disk is nevertheless still present.) Note that flop transitions of 
holomorphic disks under variation of open string moduli have been observed in 
\cite{agva,akv,wolfgang,manfred}. A transition between 4-chain and 2-chain 
domainwalls under variation of only closed string moduli was described in 
\cite{krwa}, but involved continuation through small volume phases.

\section{Outlook}

In this work, we have computed D-brane contributions to the spacetime superpotential for 
bulk fields in several compact Calabi-Yau geometries. Our B-model results are holomorphic 
invariants of the underlying quantum D-brane geometry and we have interpreted the
results in appropriate semi-classical regimes in the A-model. Among our findings
is an interesting phase transition under which the classical topology of the Lagrangian 
geometry and associated domainwalls changes. Among possible future directions, let
us mention the following three.

First of all, it will be interesting to see whether the vacuum structure in the
regime $t_2\gg t_1$ can be understood in semi-classical terms in the A-model.
This is likely to involve dynamical open string moduli, which we have suppressed
in the entire discussion.

Secondly, it would be interesting to repeat the analysis for other multi-parameter models,
to see how much of the structure survives. Residue computations on the two-parameter 
models $\projective_{11222}[8]$ and $\projective_{11669}[18]$ give results very similar 
to \eqref{verysimilar}. However, the quadratic equation \eqref{fun} is replaced by a 
quartic and cubic equation, respectively. This makes the solution of the extended 
Picard-Fuchs equations slightly more complicated.

Thirdly, one could investigate the structure of loop amplitudes in these models,
using the extended holomorphic anomaly equations of \cite{openbcov,tadpole}. 
This is interesting because the D-branes that we have studied here
presumably do not arise as fixed point sets of anti-holomorphic involutions in
the A-model. As a result, they should offer a greater degree of flexibility in 
implementing the topological tadpole cancellation condition of \cite{tadpole}.

\begin{acknowledgments}
I would like to thank 
Manfred Herbst,
Wolfgang Lerche,
David Morrison, and
Edward Witten 
for valuable discussions and comments.
\end{acknowledgments}





\end{document}